\documentclass[11pt]{article}
\usepackage[margin=1in]{geometry}
\usepackage[usenames,dvipsnames]{xcolor}
\usepackage{dsfont}
\usepackage{amsbsy}
\usepackage{amssymb, commath}
\usepackage{amsmath, mathtools}%
\usepackage{amsthm}
\usepackage{breqn}
\usepackage{graphicx}
\usepackage{wrapfig}
\usepackage{colonequals}
\usepackage{psfrag}
\usepackage{mathrsfs}
\usepackage{epstopdf}
\usepackage{subcaption}
\usepackage{url}
\usepackage{breakurl}
\usepackage{cite}

\usepackage{xr} 
\usepackage{comment}
\usepackage{diagbox}
\usepackage[toc,page]{appendix}
\usepackage{booktabs}
\usepackage{multirow}
\usepackage{bm}

\newtheorem{lem}{Lemma}

\makeatletter
\newtheorem*{rep@theorem}{\rep@title}
\newcommand{\newreptheorem}[2]{%
\newenvironment{rep#1}[1]{%
 \def\rep@title{#2 \ref{##1}}%
 \begin{rep@theorem}}%
 {\end{rep@theorem}}}
\makeatother
\newreptheorem{conj}{Conjecture}

\theoremstyle{definition}

\newtheorem{defn}{Definition}

\newcommand{\calX}{\mathcal{X}}

\newcommand{\calA}{\mathcal{A}}

\newcommand{\calM}{\mathcal{M}}

\newcommand{\calS}{\mathcal{S}}

\newcommand{\calG}{\mathcal{G}}

\renewcommand{\tilde}{\widetilde}

\DeclareMathOperator*{\argmax}{\arg\!\max}

\newcommand{\Reals}{\mathbb{R}}

\newcommand{\defined}{\triangleq}

\newcommand{\ExpVal}[2]{\mathbb{E}_{#1}\left[ #2 \right]}

\newcommand{\eps}{\varepsilon}

\newcommand{\EE}[1]{\ExpVal{}{#1}}

\newcommand{\calN}{\mathcal{N}}

\usepackage[colorlinks=true, breaklinks]{hyperref}
\hypersetup{%
,urlcolor=black
,citecolor=blue
,linkcolor=red
}

\definecolor{light-gray}{gray}{.90}

\makeatletter
\newcommand*{\addFileDependency}[1]{
  \typeout{(#1)}
  \@addtofilelist{#1}
  \IfFileExists{#1}{}{\typeout{No file #1.}}
}
\makeatother

\newtheorem{theorem}{Theorem}

\def\calA{{\mathcal A}}
\def\calS{{\mathcal S}}
\def\calM{{\mathcal M}}
\def\calX{{\mathcal X}}
\def\sE{{\mathsf E}}
\def\eps{{\varepsilon}}

\begin{document}

%

%

\title{Obfuscation via Information Density Estimation}
\date{}
\author{
    Hsiang~Hsu,
    Shahab~Asoodeh,
    and~Flavio~du~Pin~Calmon\thanks{H. Hsu, S. Asoodeh and F. P. Calmon are with the John A. Paulson School of Engineering and Applied Sciences, Harvard University, Cambridge, MA, 02138. E-mails: \texttt{hsianghsu@g.harvard.edu, shahab@seas.harvard.edu, flavio@seas.harvard.edu}.}
}
\maketitle


\begin{abstract}
Identifying features that leak information about sensitive attributes is a key challenge in the design of information obfuscation mechanisms. In this paper, we propose a framework to identify information-leaking features via information density estimation. Here, features whose information densities exceed a pre-defined threshold are deemed information-leaking features. Once these features are identified, we sequentially pass them through a targeted obfuscation mechanism with a provable leakage guarantee in terms of  $\mathsf{E}_\gamma$-divergence. The core of this mechanism relies on a data-driven estimate of the trimmed information density for which we propose a novel estimator, named the \textit{trimmed information density estimator} (TIDE). We then use TIDE to implement our mechanism on three real-world datasets. Our approach can be used as a data-driven pipeline for designing obfuscation mechanisms targeting specific features. 
\end{abstract}

\section{Introduction}
A challenging problem in dataset and information sharing platforms is limiting the leakage of sensitive or private information. Sensitive information leakage can be controlled by \emph{obfuscating} samples in a dataset prior to disclosure; i.e., perturbing the sample in a way that sensitive information cannot be effectively inferred \cite{bertran2019adversarially, chen2019distributed, huang2018generative, zemel2013learning}. Samples may contain several \textit{features}, only some of which might leak information about sensitive attributes.
%
For example, not all areas in a facial image equally disclose emotion (as a sensitive attribute), and not all terms used in Tweets equally reveal a user's political preference.
Given a  set of sensitive attributes, an information obfuscation mechanism should ideally target only those features of the data that leak excessive amount of sensitive information. 
Such mechanisms usually achieve higher utility (e.g., the quality of the image) by incorporating either complete (cf.\ information-theoretic privacy \cite{du2012privacy, Asoodeh_Arimoto, issa2018operational, hsu2018generalizing, diaz2018robustness, basciftci2016privacy}) or partial (cf.\ generative adversarial privacy \cite{huang2017context, huang2018generative}) knowledge of the underlying data distribution.

In this paper, we propose a data-driven information-obfuscation mechanism. As a natural first step, we \textit{identify} the information-leaking features in the data via an information-theoretic quantity called the \textit{information density} \cite{pinsker, han1993approximation}. 
This quantity is at the heart of most information-theoretic measures of privacy \cite{Asoodeh_Arimoto, issa2018operational,hsu2018generalizing} as well as differential privacy (DP) \cite{evfimievski2003limiting, ZeroDP, Concentrated_Dwork, balle2018improving, sarwate2013signal, chaudhuri2011differentially}.
Intuitively, the information density captures the change of the belief about a sensitive attribute upon an observation of a sample in a disclosed dataset. 
%

Features whose information density are above a certain threshold (which we call information-leaking features) can be randomized (e.g., perturbed) via an obfuscation mechanism. The goal of the obfuscation mechanism is to limit unwanted inferences about a sensitive attribute from disclosed data. We argue that this objective can be mathematically formulated in terms of a specific type of $f$-divergence \cite{Csiszar67}, called the $\sE_\gamma$-divergence, which captures the tail distribution of the information density. We propose a feature-dependent Gaussian mechanism that ensures obfuscation in terms of $\sE_\gamma$-divergence by targeting only the information-leaking features. 

The methodology proposed here aims to develop a theoretical foundation for expounding  existing approaches that completely rely on neural networks to identify and obfuscate the information-leaking features \cite{bertran2019adversarially, chen2019distributed, huang2018generative}. 
Despite its theoretical nature, our approach has a comparable performance in terms of sensitive information leakage as \cite{bertran2019adversarially}, without a specific ``utility'' target having to be pre-determined by a user. Furthermore, it adds a layer of interpretability, enabling features that pose an  excessive leakage risk to be identified and communicated to the data owner.



In practice, we need to estimate the information density from samples. This estimation problem is inherently connected to  mutual information estimation (since the expected value of information density is equal to the mutual information) which is  known to be challenging \cite{valiant2011estimating, wu2016minimax, gao2017estimating} unless an adequate parametric model is assumed \cite{vapnik2013nature}.
The main difficulty lies in the unboundedness of the information density, which leads to high sample complexity for reliable estimation.
However, since our mechanism perturbs only information-leaking features,  it requires the \emph{trimmed information density} whose estimation is a much easier task than the original information density estimation problem.
Inspired by \cite{belghazi2018mine, liu2017trimmed}, we develop the trimmed information density estimator (TIDE), based on the variational representations of $f$-divergences \cite{nguyen2010estimating, belghazi2018mine}.

The contributions of this paper, from theoretical results to practice, are listed as follows:
\begin{enumerate}
\item We propose a framework for identifying information-leaking features in terms of the trimmed information density, and use the $\sE_\gamma$-divergence between the distributions over a sensitive attribute prior and posterior to observing a disclosed sample to measure the information leakage.
Moreover, we demonstrate that obfuscation mechanisms that aim to minimize the $\sE_\gamma$-divergence 
satisfies several desirable properties in terms of information leakage guarantees (cf.\ Section~\ref{sec:privacy}).   
\item We  propose an estimator for the trimmed (thresholded) information density, named TIDE, and derive accompanying consistency and sample complexity guarantees. On the practical side, we present a neural network-based implementation for the TIDE (cf.\ Section~\ref{sec:estimation}).
\item  We apply the obfuscation mechanism in Section~\ref{sec:privacy} for  image obfuscation \cite{mcpherson2016defeating, oh2017adversarial, wu2018towards} with GENKI-$4$k \cite{genki} and Celebrity Attributes (CelebA) \cite{liu2015faceattributes} datasets, and for identifying politically-charged terms in Tweets collected from online media \cite{tweet_political_bias} (cf.\ Section~\ref{sec:exp}).
These experiments provide evidence that the TIDE can potentially serve as a building block in the design of obfuscation mechanisms.
\end{enumerate}

It is worth mentioning that information obfuscation, being inherently prior-dependent, has several limitations \cite{huang2017context}. 
In Section~\ref{sec:final_remark}, we list some of these limitations and together with our final remarks. 
Proofs, experimental details, and additional experiments on  synthetic data are provided in the Supplementary Material. 
Source code for the experiments will be made publicly available after review.

\subsection{Related Work}
The problem of balancing the competing objectives of providing meaningful information and inference from  disclosed data, on the one hand, and obfuscating sensitive information, on the other hand, has been widely studied in information-theoretic privacy \cite{du2012privacy, Asoodeh_Arimoto, issa2018operational, hsu2018generalizing, diaz2018robustness, liao2018tunable, sankar2013utility}.
Following the information-theoretic trend, these works exploit average measures (in particular mutual information and its variants) to obfuscate data. 
Recently, information obfuscation has been achieved using neural networks.
For example, in \cite{bertran2019adversarially}, an optimization problem similar to the \emph{privacy funnel} \cite{makhdoumi2014information} is formulated to train a neural network to automatically obfuscate sensitive information while maintaining utility.
In \cite{chen2019distributed, huang2017context, huang2018generative}, neural generative models are introduced  to generate ``privatized'' data that resemble the original data.
These works rely on neural networks to select and perturb features, while our approach is different in the sense that we first identify the information-leaking features using the information density and apply local obfuscation only on these features. 


Our approach of first identifying the information-leaking features and then perturbing those features is inspired by the instance-based additive mechanism of \cite{nissim2007smooth} in the DP setting.
In fact, the information density appears in DP under the name of \textit{privacy loss} variable \cite{ZeroDP, Concentrated_Dwork, balle2018improving, sarwate2013signal, chaudhuri2011differentially}, thereby connecting DP and information-theoretic quantities, e.g. mutual information DP \cite{cuff2016differential} and R\'enyi DP \cite{mironov2017renyi}.
Despite this connection, we emphasize that our approach is fundamentally different from DP, in that we consider prior distribution on sensitive attributes and also we allow correlation among features (see, e.g., \cite{Kifer_Free_Lunch} for the limitations of DP for correlated data).    

Estimating information density from samples is connected to density ratio estimation \cite{nguyen2010estimating, liu2017trimmed, yamada2011relative} --- a  fundamental task in various applications of machine learning and statistics, including outlier detection \cite{smola2009relative}, transfer learning \cite{sugiyama2007covariate}, and generative adversarial networks  \cite{goodfellow2014generative}.
A na\"ive approach to determine the density ratio is to use the plug-in estimator, which is known to perform poorly  \cite{vapnik2013nature} unless adequate parametric models (e.g., linear \cite{yamada2011relative}, kernel \cite{sugiyama2012density}, or exponential family \cite{liu2017trimmed} models) are assumed.
The two closest approaches to the trimmed information density estimation in this paper are (i)  \cite{nguyen2010estimating}, which proposed using the variational representation of $f$-divergences to convert information density estimation into an optimization problem over finite-complexity set of functions and (ii) \cite{liu2017trimmed}, which estimated the trimmed density ratio of variables from exponential family distributions.
We enforce a threshold on the information density when solving the optimization problem in the variational representation of $f$-divergences (see Section~\ref{sec:estimation}).

\subsection{Notation}
Capital letters (e.g., $X$) denote random variables, and calligraphic letters (e.g., $\calX$) denote sets.
We denote the probability measure of $X\times S$ by $P_{X, S}$, the conditional probability measure of $S$ given $X$ by $P_{S|X}$, and the marginal probability measure of $X$ and $S$ by $P_X$ and $P_S$ respectively. We use $P_{S|X}(\cdot|x)$ and $P_{S|x}$ interchangeably. 
We represent the fact that $X$ is distributed according to $P_X$ by $X \sim P_X$.
KL-divergence is given by $D_{\mathsf{KL}}(P_{S, X}\|P_SP_X) = \mathbb{E}_{P_{S, X}}[\log(P_{S, X}/P_SP_X)]$.
We denote the realization (i.e., sample) drawn from a probability distribution by $x = (x_1, \cdots, x_j, \cdots, x_m)$, where $x_{j}$ is the $j^{\text{th}}$ feature for $j = 1, \cdots, m$.
Similarly, $X_j$ is the $j^{\text{th}}$ feature of the data variable.
We denote $[k] = [1, \cdots, k]$, $x^k = [x_1, \cdots, x_k]$, and $(z)_+ = \max\{z, 0\}$ for a scalar $z$.
Finally, $\mathrm{I}_{d\times d}$ is the identity matrix of dimension $d$, and $\mathbf{1}_{\{\cdot\}}$ is the indicator function.


\section{Problem Formulation}\label{sec:privacy}
We consider the setting where a user wishes to disclose data $X$ (e.g., image, tweet)  while controlling the information revealed about a (correlated) sensitive attribute $S$ (e.g., emotion, political preference).
The goal is to produce an obfuscated representation $Y$ of $X$ that discloses only negligible information about $S$. We assume that $X$ consists of $m$ features, i.e., $X=(X_1, \dots, X_m)$, where each feature takes values in a compact set $\calX$. Throughout this section, we assume that $(S, X)\sim P_{S, X}$ and the underlying distribution $P_{S, X}$ is given. This restrictive assumption will be dropped in the subsequent section.  

One possible approach to obfuscate $X$ is to independently perturb each feature  (e.g., by adding noise to each pixel of an image). However, in many applications, only a few features of the data are correlated with the sensitive attribute,  rendering adding independent noise highly sub-optimal. 
In this section, we propose an information-theoretic framework for  data  obfuscation which consists of two parts: First, we \textit{identify}  \textit{information-leaking features}, and then obfuscate \emph{only} those features. This way, many features need not be perturbed, leading to an improvement in the utility of the disclosed data.  

Our framework relies on an information-theoretic  quantity  called  the \emph{information density}, a term coined in \cite{pinsker} and has since been used in numerous applications in information theory and statistics, particularly in binary hypothesis testing (see, e.g., Neyman-Pearson Lemma \cite{cover2012elements}). 
\begin{defn}[Information Density]
Given a pair of realization $(s, x)$ of $(S, X)\sim P_{S, X}$, the information density between $s$ and $x$ is defined as 
\begin{eqnarray}
i(s; x) \triangleq \log\frac{P_{S, X}(s,x)}{P_S(s)P_X(x)} = \log\frac{P_{X|S}(x|s)}{P_X(x)}.
\end{eqnarray}
Similarly, information density can be defined for each feature $x_j$ as
\begin{eqnarray}
i(s; x_j)\triangleq \log\frac{P_{S, X_j}(s,x_j)}{P_S(s)P_{X_j}(x_j)},
\end{eqnarray}
and the conditional information density between  $s$ and $x_j$ given another feature $x_r$ as
\begin{eqnarray}
i(s; x_j|x_r)\triangleq \log\frac{P_{S, X_j|X_r}(s,x_j|x_r)}{P_{S|X_r}(s|x_r)P_{X_j|X_r}(x_j|x_r)}.
\end{eqnarray}
\end{defn}

Intuitively, $i(s; x_j)$ evaluates the change of belief about $s$ upon observing $x_j$. 
In particular, if $|i(s; x_j)|$ is small, then $x_j$ does not significantly contribute in increasing the belief of an adversary about $s$, since $P_{S|X}(s|x_j)\approx P_{S}(s)$.  This, however, does not mean that $x_j$ can be disclosed ``as is'' without incurring an information leakage risk. 
To see why, consider, for example, that $m=2$,  $X_1$ and $X_2$ are independent and uniform binary random variables, and $S=X_1+X_2$ (modulo 2). Although $i(s; x_1)= i(s; x_2)=0$ for any realization $(s, x_1, x_2)$ of $(S, X_1, X_2)$, the release of both $x_1$ and $x_2$  would allow perfect reconstruction of $s$. To account for such inferences of sensitive attributes, we consider the conditional information density as a yardstick for identifying information-leaking features. 
\begin{defn}[Information-Leaking Feature]\label{def:information_leaking}
Given an observed sample $x=(x_1, \cdots, x_m)$, $j\in [m]$, and $\eps\geq 0$, the feature $x_j$ is said to be an \emph{$\eps$-information-leaking feature} if there exists a sensitive attribute $s$ such that $|i(s; x_j|x^{j-1})|>\eps.$
\end{defn}
The threshold $\eps$ is a tradeoff parameter between information leakage risk and the utility of the disclosed data (e.g., the quality of an image). Notice that if the data is not equipped with a natural ordering (e.g., time series), we can choose an arbitrary ordering for the conditioning features $x^{j-1}$ (cf.  Section~\ref{sec:exp_image_obfuscation} for an example in images).

\subsection{A Na\"ive Obfuscation Mechanism}
Given any $j\in [m]$, $\eps\geq 0$, and all features $x^{j-1}$, define  
\begin{eqnarray}
\begin{aligned}\label{eq:define_bad_set}
& B_j^{\eps}(x^{j-1}) \triangleq \{x\in\calX: |i(s; x_j|x^{j-1})|>\eps ~\text{for some}~s\in\calS\}.
\end{aligned}
\end{eqnarray}
If $x_j\notin B_j^{\eps}(x^{j-1})$, then it can be disclosed ``as is'' because it cannot be used to infer sensitive attributes given all the previous features.  
On the other hand, each feature $x_j\in B_j^{\eps}(x^{j-1})$ is required to be obfuscated. To do so, we shall pass all such features sequentially through an  \textit{obfuscation mechanisms} to ensure that they no longer belong to $B_j^{\eps}(x^{j-1})$.

Consider the mechanisms $\calM_j:\mathcal X\to \calX$ such that
if $x_j\notin B_j^{\eps}(x^{j-1})$ then $\calM_j(x_j) = x_j$ (deterministic)  and if $x_j\in B_j^{\eps}(x^{j-1})$ then $\calM_j(x_j)$ generates $Y_j$ a random variable from a distribution to be designed. A natural question raised here is: how should information obfuscation be measured? To answer this question, we introduce the \textit{obfuscation metric} $\Pr(|i(s; Y_j|y^{j-1}|)> \eps)$ and require    
\begin{equation}\label{constraint}
    \Pr(|i(s; Y_j|y^{j-1}|)> \eps)\leq \frac{\delta}{m},
\end{equation}
for all $s\in \calS$, 
where $y^{j-1}$ is any output of the $\calM_{1}(x_{1}), \dots, \calM_{j-1}(x_{j-1})$. Although this metric is intuitive, it presents a serious drawback for use in practice. Any reasonable mechanism must be immune to post-processing: any processing of the mechanism's output should only decrease the information leakage risk or equivalently the obfuscation metric. However, the obfuscation metric in \eqref{constraint} may violate this property.  
To see this, let $m=1$ and $\tilde Y$ be obtained by applying an arbitrary post-processing to $Y$ the output of the mechanism $\calM_1$ satisfying the obfuscation metric $\Pr(i(Y;s)>\eps)\leq \delta$ for all $s$. Immunity to post-processing is then equivalent to requiring 
\begin{equation}\label{Cotradiction}
    \Pr(i(s; \tilde Y)>\eps)\leq \Pr(i(s; Y)> \eps),
\end{equation} for all $s$, $\eps\geq 0$ and $\delta\in [0,1]$. However, we show in the following that there must exist some $\eps$ for which \eqref{Cotradiction} is violated. To see this, notice that 
$\EE{\frac{P_{\tilde{Y}|S}(\tilde{Y}|s)}{P_{\tilde{Y}}(\tilde{Y})}} =\EE{\frac{P_{Y|S}(Y|s)}{P_{Y}(Y)}} = 1$ and hence we have 
\begin{eqnarray}
\int_{0}^\infty \Pr(e^{i(s; \tilde{Y})}\geq t) \text{d}t = \int_{0}^\infty \Pr(e^{i(s; Y)}\geq t) \text{d}t.
\end{eqnarray}
Therefore, Eq.~\eqref{Cotradiction} must hold with equality for all $\eps\geq 0$ which in turn implies
\begin{eqnarray}\label{eq:counter_example}
D_{\mathsf{KL}}(P_{\tilde{Y}|s}\|P_{\tilde{Y}}) = D_{\mathsf{KL}}(P_{Y|s}\|P_{Y}).
\end{eqnarray}
However, according to data processing inequality for KL divergence, Eq.~\eqref{eq:counter_example} cannot hold true in general. Therefore, there must exist some $\eps$ for which  \eqref{Cotradiction} does not hold.
For more details about this construction, see \cite{EGamma}.

Next, we propose another metric in terms of a certain $f$-divergence, the so-called $\sE_\gamma$-divergence,  and show that it implies \eqref{constraint} while being immune to post-processing.

\subsection{$\sE_\gamma$-Divergence}
To address the issue raised above, we resort to a particular divergence metric between two probability distributions called $\sE_\gamma$-divergence, and show that this divergence bounds an appropriately weighted tail distributions of $i(s; Y)$.

\begin{defn}[$\sE_\gamma$-Divergence {\cite{polyanskiy2010channel}}]\label{Def_HS_Divergence} 
Given two probability distributions $P$ and $Q$ defined on the same support set $\mathcal A$ and $\gamma\geq 1$, we define $\sE_\gamma$-divergence as
\begin{eqnarray}
\sE_{\gamma}(P\|Q) &\triangleq& \sup_{A\subset \calA} P(A) - \gamma Q(A)\label{eq:def_e_gamma_1}\\
&=& \int_{a\in \calA}(\text{d}P(a)-\gamma \text{d}Q(a))_+\label{eq:def_e_gamma_2}, 
\end{eqnarray}
where the equality comes from  the fact that the optimizer in \eqref{eq:def_e_gamma_1} is $\calA^* = \{a \in \calA| P(a)-\gamma Q(a) \geq 0\}$.
\end{defn}

$\sE_\gamma$-divergence has been considered in various fields; for example, it appears in DP literature as an equivalent definition for differentially private mechanisms (see e.g, \cite{Barthe, Balle2019mixing}), in statistics as the probability of correct decision in Bayesian binary hypothesis testing \cite{polyanskiy2010channel}, and in information theory for proving general channel coding converse results \cite{polyanskiy2010channel, Polyanskiy_Arimoto}. 

Notice that $\sE_\gamma(P\|Q)\leq 1$ for all $\gamma\geq 1$ and any pair of distributions $(P,Q)$. It is clear that the constraint   
$\sE_{\gamma}(P_{Y}\|P_{Y|s})\leq \delta$ for some $\delta\in (0,1)$ ensures that  $P_Y(A) - \gamma P_{Y|s}(A)\leq \delta$ for \textit{all} subsets $A\subset \calX$ and in particular $P_Y(\calA^*)\leq \delta$. Since for $\gamma=e^\eps$, the set $\calA^*$ corresponds to the tail events of the random variable $i(Y;s)$, we henceforth assume $\gamma = e^\eps$.  Note also that to have control on both tail events $\{i(Y;s)<-\eps\}$ and $\{i(Y;s)>\eps\}$, we need to consider both $\sE_{e^\eps}(P_{Y}\|P_{Y|s})\leq \delta$ and $\sE_{e^\eps}(P_{Y|s}\|P_{Y})\leq \delta$. In the sequel, we present our results only for $\sE_{e^\eps}(P_{Y|s}\|P_{Y})\leq \delta$. The results for the reversed divergence can be derived \textit{mutatis mutandis}.  

Having this divergence at our disposal, we can now propose obfuscation criteria for the mechanisms $\{\calM_j\}$. As before, if $x_j\notin B^{\eps}_j(x^{j-1})$, we set $\calM_j(x_j) = x_j$; otherwise, we shall construct randomized mechanism $\calM_j:\calX\to \calX$ such that $\calM_j(x_j) = Y_j$ satisfies 
\begin{equation}\label{constraint2}
    \sE_{e^\eps}(P_{Y_j|s,y^{j-1}}\|P_{Y_j|y^{j-1}})\leq \frac{\delta}{m},
\end{equation}
where $y^{j-1}$ is a realizations of all previous mechanisms $\calM_1(x_1), \dots, \calM_{j-1}(x_{j-1})$. The factor $\frac{1}{m}$ in the right-hand side of \eqref{constraint2} is only for the sake of normalization (to be clarified in Theorem~\ref{thm:privacy_guarantee}). 

It is clear from \eqref{eq:def_e_gamma_1} that upper bounds on $\sE_\gamma(P_{Y_j|s,y^{j-1}}\|P_{Y_j|y^{j-1}})$ directly translate into low-leakage guarantee \eqref{constraint}.  Furthermore, since $\sE_\gamma$-divergence belongs to the family of $f$-divergences \cite{sason2016f}, it satisfies the data processing inequality which in turn implies that mechanisms satisfying \eqref{constraint2} are immune to post-processing.
   
To even further justify the choice of $\sE_\gamma$-divergence as a ``proxy'' for the  obfuscation metric in \eqref{constraint}, we prove  in the following theorem an equivalent formula for $\sE_{e^\eps}(P_{Y_j|s,y^{j-1}}\|P_{Y_j|y^{j-1}})$ in terms of the tail distribution $\Pr(i(s; Y_j|y^{j-1})>t)$ for $t\geq 0$.  
\begin{theorem}[Tail Distribution Formula]\label{thm:integral}
Given distributions $P_{Y_j|s,y^{j-1}}$ and $P_{Y_j|y^{j-1}}$, we have  
\begin{eqnarray}
\begin{aligned}
&\sE_{e^\eps}(P_{Y_j|s,y^{j-1}}\|P_{Y_j|y^{j-1}}) = e^{\eps}\int_{\eps}^\infty e^{-t}\Pr(i(s; Y_j|y^{j-1})>t)\text{d}t.
\end{aligned}
\end{eqnarray}
\end{theorem}
This result provides an operational interpretation for $\sE_\gamma$-divergence for our obfuscation setting. More precisely,  $\sE_{e^\eps}(P_{Y_j|s,y^{j-1}}\|P_{Y_j|y^{j-1}})\leq \delta$ enforces the events $\{i(Y;s)>t\}$ to have small aggregate (weighted) probability for \textit{all} $t\geq \eps$.

Next, we address the  composition property of the above mechanisms: If each mechanism $\calM_j$ satisfies \eqref{constraint2}, then so does the composed mechanism $\calM=(\calM_1, \dots, \calM_m)$ with parameters $m\eps$ and $\delta$. Recall that $Y=(Y_1, \dots, Y_m)$ is the output of the mechanism $\calM$.

\begin{theorem}[Composition]\label{thm:privacy_guarantee}
For all mechanisms $\calM_j, j\in[m]$ satisfying \eqref{constraint2}, we have for all $s\in \calS$ 
\begin{equation}
\sE_{e^{m\eps}}(P_{Y|s}\|P_{Y})\leq \delta.    
\end{equation}
\end{theorem}
This theorem states that a guarantee for each feature, given by \eqref{constraint2}, will result in a meaningful guarantee for the whole sample. This, in particular, demonstrates the need for conditional information density in Definition~\ref{def:information_leaking}, as opposed to the unconditional one.

\subsection{A Gaussian Obfuscation Mechanism}\label{sec:mechanism}
We next give an explicit construction of mechanisms $\{\calM_j\}$ satisfying \eqref{constraint2}. Here, we assume that each feature $x_j\in C$ where $C$ is a compact subset of $\Reals^r$. Recall that each mechanism $\calM_j$ is required to generate $Y_j$ satisfying \eqref{constraint2}.
As a simple approach to enforce this guarantee, we propose the additive Gaussian mechanism; that is, for each given $j\in [m]$, $\eps$, and $x^{j-1}\in \calX^{j-1}$, we consider the following mechanism
\begin{equation}\label{eq:mechanism}
    Y_j= x_j + \lambda \mathbf{1}_{\{x_j\in B^\eps_j(x^{j-1})\}} N,
\end{equation}
where $N$ is an independent standard Gaussian noise $\calN(0,\mathrm{I}_{r\times r})$ and $\lambda> 0$ is determined according to the following theorem.

\begin{theorem}[Gaussian Obfuscation]\label{Theorem_variance}
The Gaussian obfuscation mechanism \eqref{eq:mechanism} satisfies \eqref{constraint2} if $\lambda$ satisfies  
\begin{eqnarray}
\theta_{e^\eps}(K, \lambda)\leq \frac{\delta}{m},
\end{eqnarray}
where $K$ is the radius of $C$, i.e.,  $K=\max_{w\in C}\|w\|$, and for any $a>0$
\begin{eqnarray}\label{eq:Q_functions}
\theta_{e^\eps}(a, \lambda) \defined  \mathsf{Q}\left(\frac{\lambda \eps}{a} - \frac{a}{2\lambda}\right) - e^\eps \mathsf{Q}\left(\frac{\lambda \eps}{a} + \frac{a}{2\lambda}\right),
\end{eqnarray}
where $\mathsf{Q}(v) = \Pr(\calN(0,1)\geq v) =\int_v^\infty\frac{1}{\sqrt{2\pi}}e^{-t^2/2}\text{d}t$.
\end{theorem}
In light of this theorem, if $\eps\approx 0$, then the noise variance $\lambda$ must be of order $O(\frac{K}{-\log (1-\frac{\delta}{m})})$. The exact value of noise variance, however, cannot be derived as there is no analytic expression for the $\mathsf{Q}$ function. 

We have thus far made the information-theoretic assumption that the underlying distribution $P_{S,X}$ is given and, consequently, the information density  is known exactly. In the following section, we propose a \textit{data-driven} estimator for information density which renders our proposed mechanism applicable to real-world datasets.



\section{Trimmed Information Density}\label{sec:estimation}
The obfuscation mechanism in Section~\ref{sec:privacy} relies on the conditional information density $i(s; x_j|x^{j-1})$ to identify the set of information-leaking.  
Notice that, since information density satisfies the chain rule, i.e.
\begin{eqnarray}\label{eq:chain_rule}
i(s; x_j|x^{j-1}) = i(s; x^j) - i(s; x^{j-1}),
\end{eqnarray}
an estimate of  $i(s; x_j|x^{j-1})$ can be constructed by estimates of $i(s; x^j)$ and $i(s; x^{j-1})$. 

In general, exact estimation of the information density is hard due to its unboundeness.
However, we do not need the exact estimation; instead, we only need to know if the absolute value of the conditional information density is larger than the threshold $\eps$ (Definition~\ref{def:information_leaking}).
In other words, estimating the \emph{trimmed} information density is sufficient for obfuscation purposes.
Moreover, the tail of the information density satisfies \cite{polyanskiy2010channel}
\begin{eqnarray}\label{eq:concentration_information_density}
\Pr\left\{ i(s; X_j) > t \right\} \leq e^{-t},\; \forall s,
\end{eqnarray}
indicating that the estimation error caused by trimming the information densities can be controlled.
In this section, we propose a consistent and scalable estimator for the \emph{trimmed information density}, called the TIDE, and show that estimating the trimmed information density can be easier than estimating the exact information density in terms of sample complexity.

\subsection{Trimmed Information Density Estimator}
TIDE is based on a variational representation of KL divergence\footnote{Other $f$-divergences \cite{Csiszar67, sason2016f} could also be used, see the Supplementary Material for more details.} known as the Donsker-Varadhan (DV) representation \cite{donsker1983asymptotic}, given by 
\begin{eqnarray}\label{eq:DV_representation}
\begin{aligned}
& D_{\mathsf{KL}}(P_{S, X}\|P_SP_X) = \sup_{g: \calS\times \calX \to \Reals} \big\{\mathbb{E}_{P_{S, X}}[g(S, X)] - \log \mathbb{E}_{P_SP_X}[e^{g(S, X)}]\big\}.
\end{aligned}
\end{eqnarray}
Recall that $D_{\mathsf{KL}}(P_{S, X}\|P_SP_X)$ is equal to the mutual information $I(S; X)$ between $S$ and $X$, which is in fact the expected information density $\mathbb{E}_{P_{S, X}}[i(S, X)]$.  
It can be shown that the maximizer $g^*$ of \eqref{eq:DV_representation} is exactly the information density, i.e.,  $g^*(s, x)=i(s; x)$.  
Hence, the problem of estimating information density is equivalent to solving the functional optimization problem \eqref{eq:DV_representation} given access to samples drawn from $P_{S,X}$. 

Since the search space in \eqref{eq:DV_representation} is unconstrained, directly solving the optimization by computing the empirical expectations would fail in general. 
One practical approach is to restrict the search space to a family $\mathcal{G}(\Theta)$ of continuous and bounded functions $g_\theta$ parameterized by $\theta$ in a compact domain $\Theta \subset \Reals^d$, where $d$ is the number of parameters. 
The new constrained optimization problem corresponds to approximating the information density by a bounded function, thus the name \textit{trimmed} information density.  

The TIDE is then given by
\begin{equation}\label{eq:lle_nn_emp}
\begin{aligned}
\hat{g}_n \triangleq &\argmax_{g_\theta \in \mathcal{G}(\Theta)} \big\{\mathbb{E}_{P_{S_n, X_n}}[g_\theta(S, X)] - \log \mathbb{E}_{P_{S_n}P_{X_n}}[e^{g_\theta(S, X)}]\big\},
\end{aligned}
\end{equation}
where $P_{S_n, X_n}$ and $P_{S_n}P_{X_n}$ denote the empirical distributions of $P_{S, X}$ and $P_SP_X$ by $n$ samples, respectively.
\subsection{Consistency and Sample Complexity}
The TIDE obtained by solving \eqref{eq:lle_nn_emp} belongs to a broader class of \emph{extremum estimators} \cite{amemiya1985advanced} of the form $\hat{a} = \argmax_{a \in \calA} \Lambda_n(a)$, where $\Lambda_n(a)$ is an objective function and $\calA$ is a parameter space. 
The consistency of extremum estimators is guaranteed by the Newey-McFadden Lemma \cite{newey1994large} (cf.\ Supplementary Material), which in turn implies the consistency of the TIDE, as stated in the following theorem. 
\begin{theorem}[Consistency]\label{thm:consistency}
If $\calG(\Theta)$ is the family of continuous and bounded functions parameterized by $\theta$ taking values in a compact domain $\Theta$, then the TIDE \eqref{eq:lle_nn_emp} is consistent, i.e., for any $\eta >0$, there exist $N > 0$ such that for all $n > N$, we have  $|\hat{g}_n(s, x)-g^*(s, x)| \leq \eta$ with probability one for all $s \in \calS$ and $x \in \calX$.
\end{theorem}

We turn our attention to deriving the sample complexity of the TIDE.
We make further assumption that functions in $\calG(\Theta)$ are Lipschitz, and use \eqref{eq:concentration_information_density} to prove the following theorem.
To avoid technical complications, we assume that $\mathbb{E}_{P_{S, X}}[g(S, X)]$ and $\mathbb{E}_{P_SP_X}[e^{g(S, X)}]$ are finite for all functions $g$ in $\calG(\Theta)$.
\begin{theorem}[Sample Complexity]\label{thm:sample_complexity}
Assume that functions in $\calG(\Theta)$ are bounded by $M$ and Lipschitz with respect to $\theta$, and $\Theta \subset \Reals^d$ is compact.
Then we have $|\hat{g}_n(s, x)-g^*(s, x)|\leq \eta$ with probability at least $1-e^{-M}$, for all $s \in \calS$ and  $x\in\calX$, where $n = O(\frac{M^3 d}{\eta^2})$.
\end{theorem}
Observe that trimming the information density is crucial for the bound in the previous theorem to hold: if $M \to \infty$ (i.e., estimating the exact information density), the sample complexity of the TIDE grows to infinity and the result is vacuous.
In fact, we need to restrict the search space to all continuous and bounded functions $\calG$ to exactly approximate the trimmed information density. However, for computational reason, we assume that these functions can be parameterized by a compact domain $\Theta$, and the complexity of the family $\calG(\Theta)$ is characterized by its number of parameters $d$.
As the complexity of the functions $d \to \infty$, meaning the search space is too large, the sample complexity goes to infinty. 
This assumption allows us to approximate the functions in $\calG(\Theta)$ by neural networks, where $\Theta$ is the weights in all layers, as we will see next.


\subsection{Implementation}\label{sec:implementation}
In practice, we use the set of functions representable by a neural network with output clipped to $[-M,M]$ to approximate the set of continuous and bounded functions $g(s, x)$ in $\calG$. 
By sampling $(s, x)$ from $P_{S, X}$ and from $P_S\times P_X$ for the first and second expectations in \eqref{eq:lle_nn_emp}, we can fit the neural network.
After training, the $g(s, x)$ outputs the estimate of the trimmed information density of samples $|i(s; x)| \leq M$.
In order to reconstruct the conditional information density by the chain rule \eqref{eq:chain_rule}, we compute $g(s, x^j)$ for $i(s, x^j)$ and $g(s, x^{j-1})$ for $i(s, x^{j-1})$; then the $i(s, x^j)-i(s, x^{j-1})$ gives the desired conditional information density $|i(s; x_j|x^{j-1})| \leq 2M$.

\section{Experiments}\label{sec:exp}
The experiments contain two parts. 
First, we investigate image obfuscation \cite{mcpherson2016defeating, oh2017adversarial, wu2018towards} as a common use case of our approach with the GENKI-$4$k \cite{genki} and Celeberity Attributes \cite{liu2015faceattributes} datasets.
Second, we demonstrate how TIDE can be possibly used to discover politically-charged terms in the Tweets of online media \cite{tweet_political_bias}.
Detailed experimental setups (e.g., architecture of the function $g$ in TIDE, training details) and additional experiments on Gaussian synthetic data are provided in the Supplementary Material.


\subsection{Image Obfuscation}\label{sec:exp_image_obfuscation}
A common application of information obfuscation is image obfuscation \cite{mcpherson2016defeating, oh2017adversarial, wu2018towards}, where we aim to hide information related to a given sensitive attribute in an image.
Unlike existing works which rely on neural networks to select and perturb features \cite{mcpherson2016defeating, bertran2019adversarially}, we apply the TIDE to identify information-leaking features for the Gaussian obfuscation mechanism (Section~\ref{sec:privacy}).
We split $x$ into a grid, where each ``patch'' of size $p \times p$ pixels in the grid represents the low-level features $x_j$ of the image $x$.
It is a common method to extract low-level features in an image \cite{mcpherson2016defeating}.
We number each $x_j$ in an image from the upper-left corner to the lower-right, and use the TIDE (with $M = 3$) to determine the information-leaking features by \eqref{eq:define_bad_set}, and demonstrate our obfuscation approach on two datasets: the GENKI-$4$k and Celeberity Attributes datasets.

\begin{figure}[!tb]
\centering
\includegraphics[width=0.3\textwidth]{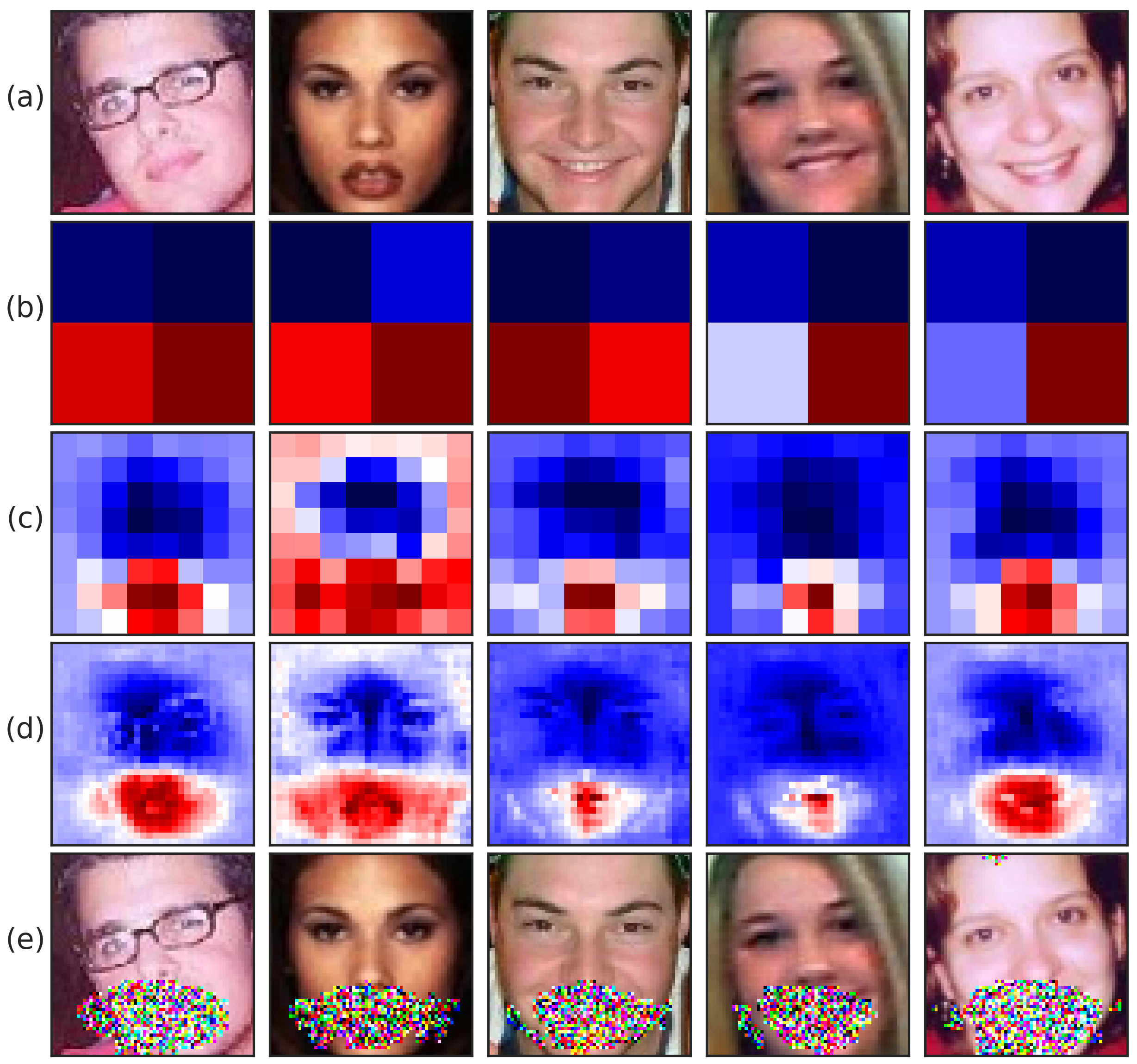}
\caption{\footnotesize 
Row (a) shows original images. Rows (b), (c) and (d) show the information-leaking patches found by the TIDE {\eqref{eq:lle_nn_emp}} with patch sizes $32\times 32$, $8\times 8$ and $2\times 2$ pixels respectively (color red indicates higher value).
Row (e) shows the Gaussian obfuscation mechanism {\eqref{eq:mechanism}} on row (d) with $\eps = 0.5$ and $\lambda = 1.0$, which successfully hide the sensitive attribute of emotion.
The information-leaking patches is easy to interpret: the TIDE focuses more on the mouth area as the patches become finer.
}
\label{fig:genki}
\end{figure}

\subsubsection{GENKI-$4$K Dataset}
This dataset contains $2400$ images for training and $600$ for testing, where each image $x$ is a $64\times 64$ pixels face that has emotion smiling ($s = 1$) or not ($s = 0$).
We select $10$ faces for illustration in Figure~\ref{fig:genki}.
When the patch size is $32\times 32$ ($4$ patches), the TIDE simply flags the lower two patches to be information-leaking. 
As the patch becomes finer, the information-leaking patches concentrate to the mouse area; thus when applying the Gaussian obfuscation mechanism, it is visually possible to identify the gender of the subject but with their emotion obfuscated. 
The leakage guarantee in Theorem~\ref{Theorem_variance}, $\delta/m \approx 0.24$, can be computed by \eqref{eq:Q_functions} with $\eps = 0.5$ and $K = 1$ since the images are normalized.
Note that the TIDE can not only reveal the patches informative of emotion, but also captures the contour of faces.

We train an adversary that can classify the emotion of the subject with accuracy $92.04\%$, and report the classification accuracy of the Gaussian obfuscation mechanism ($\lambda=1$ in \eqref{eq:mechanism}) under different patch sizes and threshold $\eps$ in Table~\ref{tab:genki_patch_vs_eps}.
When $\eps = \infty$ (i.e. $B_j^{\eps}(x^{j-1}) = \phi$ for all $j$), no patch is identified by the TIDE, and therefore the performances are the same as the adversary.
A simple mechanism to hide the emotion in images is adding Gaussian noise onto the Lower Half of the Image (LHI).
As a comparison, the results of LHI and random guessing are also included in Table~\ref{tab:genki_patch_vs_eps}.
The LHI gives similar performance when the patch size is $32 \times 32$ since when $\eps = 0.5$, the lower two patches of the image will be identified as information-leaking for the mechanism (Figure~\ref{fig:genki} row (b)), but LHI will erase too much information that is not related to the emotion.
The random guessing values correspond the to prior distribution of the emotion labels in the training set.

\begin{table}[!tb]
  \footnotesize
  \caption{\footnotesize 
  Classification accuracy of emotion obfuscation for the GENKI-$4$k dataset with different patch sizes $p\times p$ and threshold $\eps$.
  Results on obfuscating the lower-half image by Gaussian noise (LHI) and on random guessing are shown as comparison. 
  }
  \label{tab:genki_patch_vs_eps}
  \centering
  \begin{tabular}{lclclclclclcl}
     & \multicolumn{5}{c}{Classification Accuracy $\%$} \\
    \cmidrule(r){2-6}  
    \diagbox{$p\times p$}{$\eps$} & $0.5$ & $0.6$ & $0.7$ & $0.8$ & $\infty$ \\
    \hline
    \hfil $32 \times 32$ & \bm{$50.54$} & $50.54$ & $92.04$ & $92.04$ & $92.04$ \\
    \hfil $16 \times 16$ & \bm{$50.72$} & $51.46$ & $79.14$ & $89.52$ & $92.04$ \\
    \hfil $8 \times 8$   & \bm{$50.93$} & $68.94$ & $78.71$ & $87.33$ & $92.04$ \\
    \hfil $4 \times 4$   & \bm{$50.60$} & $65.06$ & $75.23$ & $83.89$ & $92.04$ \\
    \hfil $2 \times 2$   & \bm{$50.64$} & $62.25$ & $68.59$ & $80.26$ & $92.04$ \\
    \hline
    \hfil LHI   & \bm{$50.58$} & \hfil - & \hfil - & \hfil - & \hfil -\\
    \hfil Guess & \bm{$50.41$} & \hfil - & \hfil - & \hfil - & \hfil -\\
  \end{tabular}
\end{table}

\subsubsection{Celebrity Attributes (CelebA) Dataset}
\begin{figure}[!tb]
\centering
\includegraphics[width=0.3\textwidth]{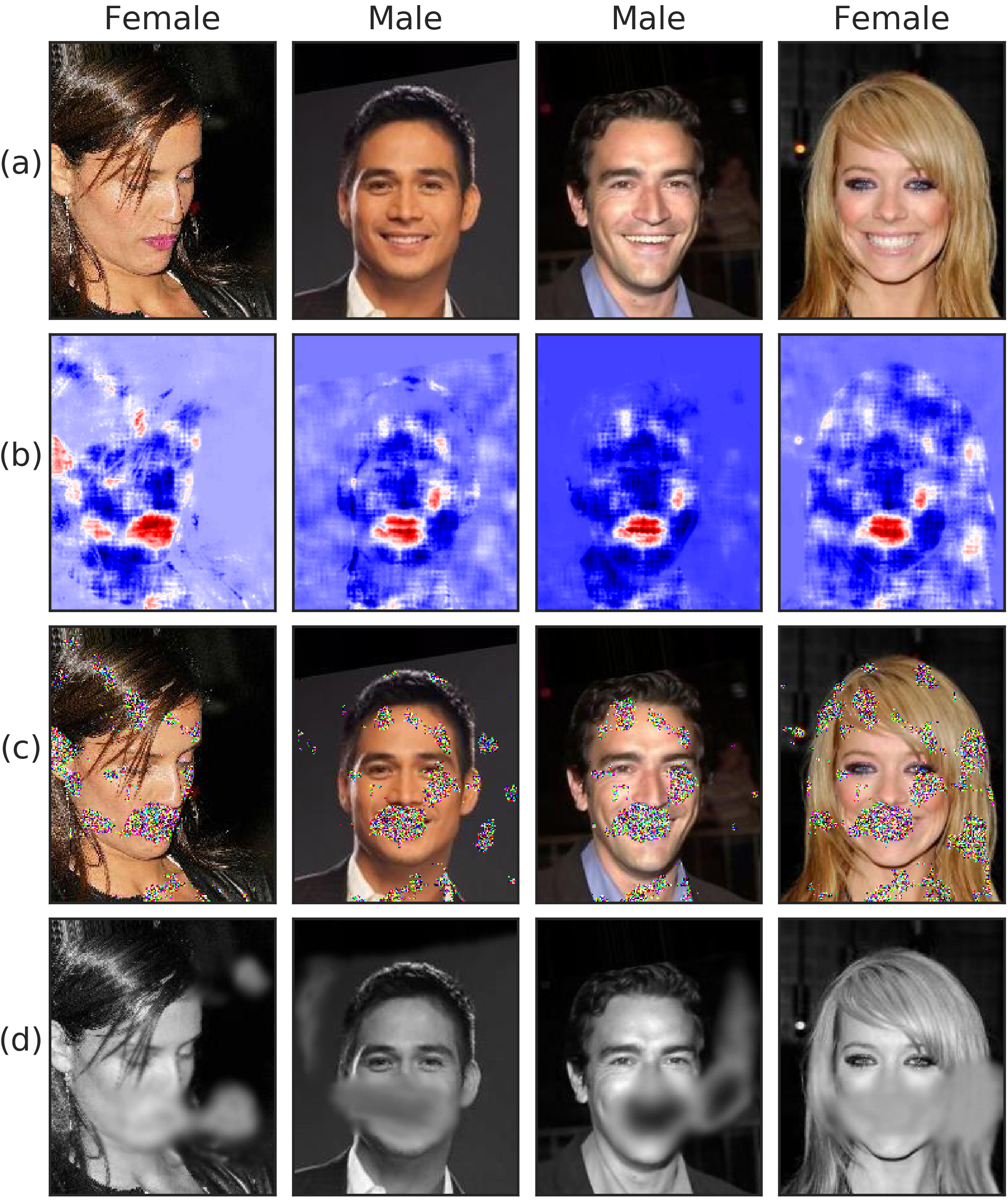}
\caption{\footnotesize 
Row (a) shows original images. Row (b) shows the information-leaking patches with size $2\times 2$ by the TIDE (color red indicates higher value).
Row (c) shows the Gaussian obfuscation mechanism on row (b) with $\eps = 0.74$ and $\lambda = 1$, and row (d) shows information obfuscation in \cite{bertran2019adversarially} with the sensitive information budget equal to $0.72$ bits.
}
\label{fig:celeba}
\end{figure}

This more challenging dataset contains $202599$ colorful high-resolution images, where each image is a $218\times 178$-pixel face image of a celebrity with $40$ distinct binary labels, including \texttt{smiling}, \texttt{gender}, \texttt{Arched Eyebrows}, etc. 
We select $100$k images as $X$ and the sensitive attribute $S$ to be emotion as well for training the TIDE.
In Figure~\ref{fig:celeba}, we randomly pick $4$ images for illustration.
Given a small patch size, the Gaussian obfuscation mechanism ($\lambda=1$ in \eqref{eq:mechanism}) perturbs selective patches to hide the sensitive attribute while keeping other useful information (e.g. gender) intact.
The leakage guarantee (Theorem~\ref{Theorem_variance}), $\delta/m \approx 0.18$, can be computed by \eqref{eq:Q_functions} with $\eps = 0.74$ and $K = 1$.
In Figure~\ref{fig:celeba} row (d), we reproduce the method by \cite{bertran2019adversarially} since it is the state-of-the-art result in information obfuscation and its implementation is publicly available.
The main difference between our approach and \cite{bertran2019adversarially} is that \cite{bertran2019adversarially} requires an additionally pre-specified utility (i.e. the labels of gender), while our approach does not require such labels.
As we can see, both methods shown in Figure~\ref{fig:celeba} rows (c) and (d) obfuscate the mouth and some other area.
However, our approach tends to obfuscate less of the subject's face.

We train two classifiers for emotion and gender, and report the accuracy of our approach and \cite{bertran2019adversarially} in Table~\ref{tab:celeba}.
Both methods block emotion recognition, effectively pushing the accuracy of the emotion classifier towards random guessing.
More importantly, the gender classifier still performs well over the sanitized images.

\begin{table}[!tb]
  \footnotesize
  \caption{\footnotesize Comparison between our approach (with patch size $2\times 2$) and \cite{bertran2019adversarially} ($\epsilon$ here stands for the tolerance of sensitive information leak) on emotion and gender classification accuracy for the CelebA dataset.
  }
  \label{tab:celeba}
  \centering
  \begin{tabular}{lclclclclcl}
    & \multicolumn{4}{c}{Classification Accuracy $\%$} \\
    \cmidrule(r){2-5}
    \hfil Threshold & \multicolumn{2}{c}{Our approach} & \multicolumn{2}{c}{Method in \cite{bertran2019adversarially}} \\
    \cmidrule(r){2-3}   \cmidrule(r){4-5}
    \hfil $\eps$ & \hfil Emotion & \hfil Gender & \hfil Emotion & \hfil Gender \\
    \hline
    \hfil $\infty$ & \hfil $92.04$ & \hfil $94.29$ & \hfil $92.04$ & \hfil $94.29$\\
    \hfil $0.8$    & \hfil $85.97$ & \hfil $91.48$ & \hfil $85.59$ & \hfil $92.53$\\
    \hfil $0.7$    & \hfil $75.15$ & \hfil $90.39$ & \hfil $76.40$ & \hfil $91.20$\\
    \hfil $0.6$    & \hfil $71.33$ & \hfil $87.61$ & \hfil $70.88$ & \hfil $89.77$\\
    \hfil $0.5$    & \hfil \bm{$69.01$} & \hfil \bm{$86.97$} & \hfil $68.60$ & \hfil $89.47$\\
    \hline
    \hfil LHI   & \hfil $53.91$ & \hfil $69.35$ & \hfil $53.91$ & \hfil $69.35$\\
    \hfil Guess & \hfil $51.79$ & \hfil $58.32$ & \hfil $51.79$ & \hfil $58.32$\\
  \end{tabular}
\end{table}

\subsection{Information-Leaking Terms in Tweets}
Finally, we showcase how the TIDE can be used in natural language to identify politically-charged terms in the Tweets from online media \cite{tweet_political_bias}.
The information density is called the pointwise mutual information (PMI) in natural language processing to measure associations between words and labels \cite{church1990word}.
Since perturbation on languages is not yet well-defined \cite{alzantot2018generating}, we do not perform the mechanism in \eqref{eq:mechanism}, but focus on identifying information-leaking terms.

\begin{figure}[!tb]
\centering
\includegraphics[width=0.49\textwidth]{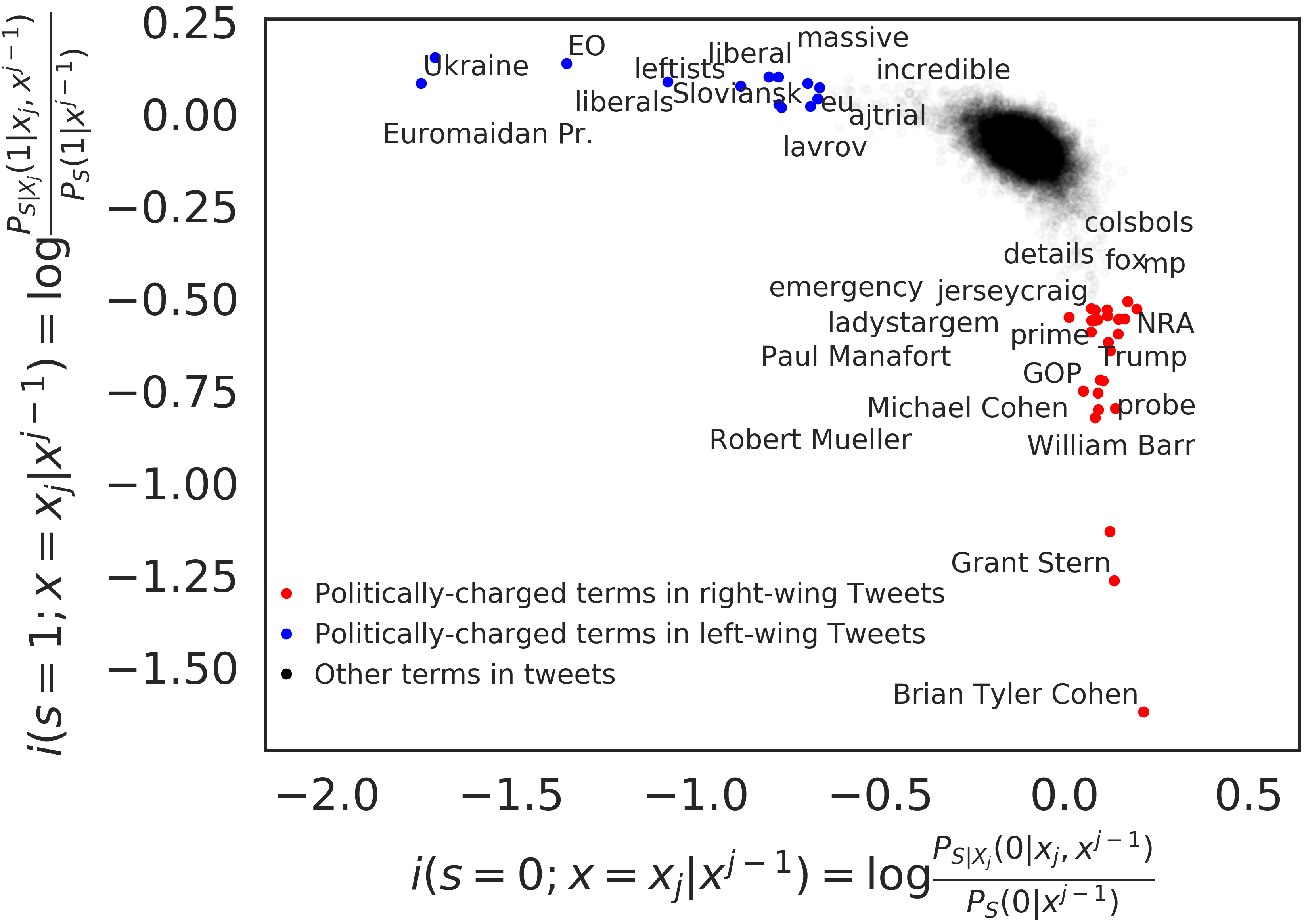}
\caption{\footnotesize
  $i(s; x_j|x^{j-1})$ for terms in Tweets.
  GOP: Grand Old Party (i.e. the Republican Party), NRA: National Rifle Association, EO: Entrepreneurs' Organization, Euromaidan Pr.: Euromaidan Press.}
\label{fig:tweet}
\end{figure}

We collect $N = 75946$ Tweets from more than $20$ online publishers (e.g. CNN, Bloomberg, New York Times), and determine their private attribute $S$ as the political preference of being right-wing ($s=0$) and left-wing ($s=1$) according to \cite{tweet_political_bias}, where the numbers of samples with each political bias are equivalent.
We pre-process the Tweets to keep only meaningful terms (i.e. pieces of words) and use bag-of-words representation \cite{manning2010introduction} to tokenize all the pieces of words for each Tweet according to term frequency, ending up with $24657$ terms (i.e. features $x_j$, $j \in [24657]$).
We train the TIDE using the tokenized Tweets as $x$.
In Figure~\ref{fig:tweet}, we show the estimate of trimmed conditional information density $i(s; x_j|x^{j-1})$ of each term. It is clear that some terms carry more information about the political bias.
For instance, terms such as ``Grand Old Party'' and ``National Rifle Association'' associate with right-wing politics, and terms ``Europe'' and ``liberal(s)'' with the left.
In this scenario, our approach could be eventually deployed as a plug-in to warn the users about potential political preference leaks before posting Tweets.

\section{Final Remarks}\label{sec:final_remark}
We introduced a new information obfuscation framework that first identifies information-leaking features using the trimmed information density, and then tailors the obfuscation mechanism only on these features.
To our knowledge, this framework is the first formal application of information density to quantify information-leaking features, and could potentially serve as a data-driven tool for designing obfuscation mechanism for high-dimensional data. 
\paragraph{Limitations.}
In order to estimate the information density, we make two key assumptions: (i) we know \emph{a priori} sensitive attributes that we wish to hide (e.g., political preference), and (ii) we have access to a reference dataset from which we can fit the TIDE (though this is difficult to avoid as discussed in \cite{vzliobaite2016using}).
Although these assumptions are restrictive in practice, they allow us to develop systematic machinery to discover information-leaking samples and features in an entirely data-driven manner.

\appendix
Here, we give proofs of theorems and other technical discussions omitted from Sections~\ref{sec:privacy} and \ref{sec:estimation} and also provide further details about the experiment setups and also the training phase. 

\section{Proofs and Theoretical Backgrounds}
In this section, we provide proofs omitted in the main text, as well as some discussions on the relationship between the TIDE and variational representations of $f$ divergences and the Newey-McFadden lemma.




\subsection{Proof of Theorem~\ref{thm:integral}}
For notational brevity, we drop $y^{j-1}$ from the conditioning part of $P_{Y_j|s, y^{j-1}}$ and $P_{Y_j|y^{j-1}}$ and also write $P$ and $Q$ for $P_{Y_j|s}$ and $P_{Y_j}$, respectively. 
To prove this theorem, note that according to Definition~\ref{Def_HS_Divergence}, we can write 
$$\mathsf{E}_{e^\eps}(P\|Q) = P(i(s; Y_j)>\eps) - e^\eps Q(i(s; Y_j)>\eps).$$
Hence, letting $\mathcal C$ denote the tail event $\{y: i(s; y)>\eps\}$ for a given $s$, we have 
\begin{eqnarray*}
    	\mathsf{E}_{e^\eps}(P\|Q)&=& P(\mathcal C) - e^\eps Q(\mathcal C)\\
    &=& \mathbb{E}_{P}\left[\mathbf{1}_{\{Y_j\in \mathcal C\}}\right] -  e^\eps\mathbb{E}_{Q}\left[\mathbf{1}_{\{Y_j\in \mathcal C\}}\right]\\
    &\stackrel{(a)}{=}& \mathbb{E}_Q\left[e^{i(s; Y_j)}\mathbf{1}_{\{Y_j\in \mathcal C\}}\right]- e^\eps\mathbb{E}_Q\left[\mathbf{1}_{\{Y_j\in \mathcal C\}}\right]\\
    &=&\mathbb{E}_Q\left[\left(e^{i(s; Y_j)}-e^\eps\right)\mathbf{1}_{\{Y_j\in\mathcal C\}}\right]\\
    &=&\mathbb{E}_Q\left[\left(e^{i(s; Y_j)}-e^\eps\right)_+\right]\\
     &=&\mathbb{E}_Q\left[e^{i(s; Y_j)}e^{-i(s; Y_j)}\left(e^{i(s; Y_j)}-e^\eps\right)_+\right]\\
    &\stackrel{(b)}{=}&\mathbb{E}_P\left[\left(1-e^\eps e^{-i(s; Y_j)}\right)_+\right]\\
    &=&\int_{0}^\infty\Pr\left(\left(1-e^{\eps-i(s; Y_j)}\right)1_{\{Y_j\in\mathcal C\}}\geq t\right)\text{d}t,
	\end{eqnarray*} 
where both $(a)$ and $(b)$ follow from the  simple change-of-variable argument $\mathbb{E}_P\left[f(Y)\right] = \mathbb{E}_Q\left[e^{i(s; Y_j)}f(Y)\right]$ for any function $f$.

Furthermore, since $\left(1-e^{\eps-i(s; Y_j)}\right)1_{\{i(s; Y_j)>\eps\}}<1$ with probability one, we have
\begin{eqnarray*}
\mathsf{E}_{e^\eps}(P\|Q)&=& \int_{0}^\infty\Pr\left(\left(1-e^{\eps-i(s; Y_j)}\right)1_{\{Y_j\in\mathcal C\}}\geq t\right)\text{d}t\\
    &=&  \int_{0}^1\Pr\left(\left(1-e^{\eps-i(s; Y_j)}\right)1_{\{Y_j\in\mathcal C\}}\geq t\right)\text{d}t\\
    &=&\int_{0}^1\Pr\left(1-e^{\eps-i(s; Y_j)}\geq t\right)\text{d}t\\
    &=&\int_{0}^1\Pr\left(e^{-i(s; Y_j)}\leq (1-t)e^{-\eps}\right)\text{d}t\\
    &=&e^\eps\int_{0}^{e^{-\eps}}\Pr\left(e^{-i(s; Y_j)}\leq b\right)\text{d}b\\
    &=&e^\eps\int_{\eps}^{\infty}e^{-t}\Pr\left(i(s; Y_j)\geq t\right)\text{d}t.
\end{eqnarray*}

\subsection{Proof of Theorem~\ref{thm:privacy_guarantee}}
First assume that $m=2$. For any set $A\subset \calX^2$ and $s\in \calS$, we have 
\begin{align*}
P_{Y_1Y_2|s}(A) &= \sum_{y_1\in \calX} P_{Y_1|s}(y_1)\Pr((y_1, Y_2)\in A|s) \\
&\leq \sum_{y_1\in \calX} P_{Y_1|s}(y_1)\min\left\{1, e^\eps\Pr((y_1, Y_2)\in A)+\delta'\right\}\\
&\leq \sum_{y_1\in \calX} P_{Y_1|s}(y_1)\min\left\{1, e^\eps\Pr((y_1, Y_2)\in A)\right\} + \delta'\\
& \leq  \sum_{y_1\in \calX} \left(e^\eps P_{Y_1}(y_1) + \zeta(y_1)\right) \min\left\{1, e^\eps\Pr((y_1, Y_2)\in A)\right\}+\delta'\\
&\leq  \sum_{y_1\in \calX} e^\eps P_{Y_1}(y_1)\min\left\{1, e^\eps\Pr((y_1, Y_2)\in A)\right\} + \sum_{y_1\in \calX}\zeta(y_1) + \delta'\\
&\leq  e^{2\eps}\sum_{y_1\in \calX} P_{Y_1}(y_1)\Pr((y_1, Y_2)\in A)+ \sum_{y_1\in \calX}\zeta(y_1) + \delta'\\
&\leq  e^{2\eps}P_{Y_1Y_2}(A) + \sum_{y_1\in \calX}\zeta(y_1) + \delta'\\
&= e^{2\eps}P_{Y_1Y_2}(A) + \sE_{e^\eps}(P_{Y_1|s}\|P_{Y_1}) + \delta'\\
&\leq  e^{2\eps}P_{Y_1Y_2}(A) + 2\delta'
\end{align*}
where $\delta'=\frac{\delta}{2}$ and $\zeta(a)\coloneqq \left(P_{Y_1|s}(a)-e^\eps P_{Y_1}(a)\right)_+$ for any $a\in \calX$. The last step follows from the fact that $\sE_\gamma(P\|Q) = \sum_{a\in \calX}(P(a)-\gamma Q(a))_+$. 
Consequently, we obtain that 
$$P_{Y_1Y_2|s}(A)\leq e^{2\eps}P_{Y_1Y_2}(A) + \delta,$$
for any set $A\subset \calX^2$ for $m=2$. Repeating this argument $(m-1)$ times, we can write 
$$P_{Y|s}(A)\leq e^{m\eps}P_{Y}(A) + \delta,$$
for any set $A\subset\calX^m$ and $s\in \calS$ from which we conclude 
$$\sE_{e^\eps}(P_{Y|s}\|P_{Y})\leq \delta.$$

\subsection{Proof of Theorem~\ref{Theorem_variance}}
For any $\gamma\geq 1$ and $y^{j-1}\in\calX^{j-1}$, we have 
\begin{align}
\sE_\gamma(P_{Y_j|s, y^{j-1}}\|P_{Y_j|y^{j-1}})&\leq \sup_{x^{j-1}}\sE_\gamma(P_{Y_j|s,x^{j-1}, y^{j-1}}\|P_{Y_j|x^{j-1},y^{j-1}})\nonumber\\
&=\sup_{x^{j-1}}\sE_\gamma(P_{Y_j|s,x^{j-1}}\|P_{Y_j|x^{j-1}})\label{Bounds1},
\end{align}
where the inequality follows from the convexity of $\sE_\gamma$-divergence in each of its arguments (see, e.g., \cite{sason2016f}). 
Notice that for any given $x^{j-1}\in \calX^{j-1}$, we can write (with an abuse of notation)
\begin{align*}
\sE_\gamma(P_{Y_j|s,x^{j-1}}\|P_{Y_j|x^{j-1}})&= \int_{B}[P(\text{d}x_j|s, x^{j-1})\calN(x_j, \lambda)- e^\eps P(\text{d}x_j|x^{j-1}) \calN(x_j, \lambda)]_+\\
&\qquad\qquad + \int_{B^{\mathsf{c}}}[P(\text{d}x_j|s, x^{j-1})- e^\eps P(\text{d}x_j|x^{j-1})]_+\\
&=\int_{B}[P(\text{d}x_j|s, x^{j-1})\calN(x_j, \lambda) - e^\eps P(\text{d}x_j|x^{j-1}) \calN(x_j, \lambda)]_+
\end{align*}
where we use $B$ and $B^{\mathsf{c}}$ to write $B_j^\eps(x^{j-1})$ and its complement. This demonstrates that the mass points corresponding to the event $B^{\mathsf{c}}$ do not contribute in the $\sE_\gamma$-divergence.

Letting $P = P_{X_j|s, x^{j-1}}$ and $Q = P_{X_j| x^{j-1}}$ for a given $x^{j-1}$, it follows from above that 
\begin{align}
\sup_{x^{j-1}} &~\sE_\gamma(P_{Y_j|s,x^{j-1}}\|P_{Y_j|x^{j-1}})  \leq  \sE_\gamma(P*\calN(0,\lambda)\|Q*\calN(0,\lambda))  = \EE{\sE_\gamma (\calN(A,\lambda))\|\calN(B,\lambda)}\label{Coupling},
\end{align}
where $*$ denotes the convolution operator and $A\sim P$ and $B\sim Q$  and the expectation is taken over any arbitrary coupling of $P$ and $Q$ (e.g., their product). It can be shown that 
\begin{align}
&\sE_{\gamma}(\calN(\mu_1, \lambda^2\mathrm{I}_r)\| \calN(\mu_2, \lambda^2\mathrm{I}_r))=\mathsf{Q}\left(\frac{\log\gamma}{\beta} - \frac{1}{2}\beta\right) - \gamma \mathsf{Q}\left(\frac{\log\gamma}{\beta} + \frac{1}{2}\beta\right), \label{E_Gamma_Gaussian}
\end{align}
where $\mathsf{Q}(v) = \Pr(\calN(0,1)\geq v) =\int_v^\infty\frac{1}{\sqrt{2\pi}}e^{-t^2/2}\text{d}t$ and $\beta=\frac{\|\mu_1-\mu_2\|}{\lambda}$.
Notice that the $\sE_\gamma$-divergence between two Gaussian distributions depends on their means only through their differences. 
$$\theta_\gamma(a, \lambda)\triangleq \sE_{\gamma}(\calN(\mu, \lambda^2\mathrm{I}_r)\| \calN(0, \lambda^2\mathrm{I}_r)),$$
where $\|\mu\|= a$. 
According to \eqref{Coupling}, we can now write 
$$\sup_{x^{j-1}} \sE_\gamma(P_{Y_j|s,x^{j-1}}\|P_{Y_j|x^{j-1}})\leq \sup_{a\in C}\theta_\gamma(\|a\|, \lambda) = \theta_\gamma(K, \lambda),$$
where the equality is due to the fact that $a\mapsto \theta_\gamma(a, \lambda)$ is increasing for a fixed $\lambda$. This, together with \eqref{Bounds1}, implies  
$$\sE_\gamma(P_{Y_j|s, y^{j-1}}\|P_{Y_j|y^{j-1}})\leq \theta_{\gamma}(K, \lambda),$$
and hence \eqref{eq:mechanism} is satisfied if 
$\theta_{e^\eps}(K, \lambda)\leq \frac{\delta}{m}$.

\subsection{Estimating Information Density using $f$-Divergences}\label{appendix:f_divergence}
Other $f$-divergence measures could also be used to estimate the information density by leveraging their dual representation {\cite{nguyen2010estimating}}.
Given a convex function $f$ with $f(1) = 0$, the $f$-divergence $D_f(P \| Q) = \mathbb{E}_Q f\left(\frac{P}{Q}\right)$ can be expressed as
\begin{eqnarray}
D_f(P \| Q) = \sup_{g: \calX \to \Reals} \mathbb{E}_{P}[g(X)] - \mathbb{E}_Q[f^*(g(X))],
\end{eqnarray}
where $f^*(t) \triangleq \sup_{x \in \Reals} \{ xt - f(t) \}$ is the Fenchel convex conjugate of $f$. It can be shown that the optimizer is the subdifferential $\partial f(\frac{P}{Q})$ which, in turn, is a non-decreasing function of $\frac{P}{Q}$. 
Thus, $D_f(P\|Q)$ is also a candidate loss function in  density ratio estimation problems.

\subsection{Newey-McFadden Lemma}\label{appendix:nm_lemma}
\begin{lem}[{\cite[Theorem~2.1]{newey1994large}}]\label{lem:consistency}
Given the extremum estimator $\hat{a} = \argmax_{a \in \calA} \Lambda_n(a)$, if (i) $\calA$ is compact; (ii) there exists a limiting function $\Lambda(a)$ such that $\Lambda_n(a)$ converges to $\Lambda(a)$ in probability over $\calA$; (iii) $\Lambda(a)$ is continuous and has unique maximum at $a = a^*$, then $\hat{a}$ is a consistent estimator of $a^*$.
\end{lem}

\subsection{Proof of Theorem~\ref{thm:consistency}}
Let the objective function of the extremum estimator be
\begin{eqnarray}
\Lambda_n(g) \triangleq \mathbb{E}_{P_{S_n, X_n}}[g(S, X)] - \log \mathbb{E}_{P_{S_n}P_{X_n}}[e^{g(S, X)}].
\end{eqnarray}
We prove this theorem by checking the properties of $\Lambda_n(g)$ according to Lemma~\ref{lem:consistency}.
First, since $\Theta$ is compact and the mappings $g_\theta$ are continuous, the images $\mathcal{G}(\Theta)$ is also compact. 
Second, by triangular inequality, for $g \in \mathcal{G}(\Theta)$, we have 
\begin{eqnarray}\label{eq:consisten_3}
\begin{aligned}
& |\Lambda_n(g) - (\mathbb{E}_{P_{S, X}}[g(S, X)] - \log \mathbb{E}_{p_Sp_X}[e^{g(S, X)}])|\\
\leq &\sup\limits_{g \in \mathcal{G}(\Theta)} |\mathbb{E}_{P_{S, X}}[g(S, X)] -\mathbb{E}_{P_{S_n, X_n}}[g(S, X)]|\\
& +  \sup\limits_{g \in \mathcal{G}(\Theta)} |\log\mathbb{E}_{P_{S}P_{X}}[g(S, X)] - \log\mathbb{E}_{P_{S_n} P_{X_n}}[g(S, X)]|.
\end{aligned}
\end{eqnarray}
Since the function $g$ is uniformly bounded by $M$, i.e. $|g| \leq M$ for all $\theta$, $s$ and $x$, and logarithm is Lipschitz continuous with constant $e^M$ in the interval $[e^{-M}, e^M]$, we have
\begin{eqnarray}\label{eq:consisten_4}
\begin{aligned}
&|\log\mathbb{E}_{P_{S}P_{X}}[g(S, X)] - \log\mathbb{E}_{P_{S_n} P_{X_n}}[g(S, X)]|\leq e^M |\mathbb{E}_{P_{S}P_{X}}[g(S, X)] - \mathbb{E}_{P_{S_n} P_{X_n}}[g(S, X)]|.
\end{aligned}
\end{eqnarray}
Moreover, since $\mathcal{G}$ is compact and $g$ is continuous, the functions $g$ and $e^g$ satisfy the uniform law of large numbers \cite{van2000empirical}. Thus, Given $\eta > 0$, there exists an integer $N$ such that for all $n \geq N$ and with probability one, 
\begin{eqnarray}\label{eq:consisten_5}
\sup\limits_{g \in \mathcal{G}(\Theta)} |\mathbb{E}_{P_{S, X}}[g(S, X)]-\mathbb{E}_{P_{S_n, X_n}}[g(S, X)]| \leq \frac{\eta}{2},
\end{eqnarray}
and
\begin{eqnarray}\label{eq:consisten_6}
&&\sup\limits_{g \in \mathcal{G}(\Theta)} |\log\mathbb{E}_{P_{S}P_{X}}[g(S, X)] - \log\mathbb{E}_{P_{S_n} P_{X_n}}[g(S, X)]|\leq \frac{\eta}{2}e^{-M}.
\end{eqnarray}
Summarizing (\ref{eq:consisten_3})-(\ref{eq:consisten_6}), we have with probability one
\begin{eqnarray}
&& |\Lambda_n(g) - (\mathbb{E}_{P_{S, X}}[g(S, X)] - \log \mathbb{E}_{P_SP_X}[e^{g(S, X)}])| \leq \eta.
\end{eqnarray}
In other words, there exists a limiting function $\Lambda(g) = \mathbb{E}_{P_{S, X}}[g(S, X)] - \log \mathbb{E}_{P_SP_X}[e^{g(S, X)}]$ such that $\Lambda_n(g)$ converges to $\Lambda(g)$ in probability.

Third, since $\Lambda(g) = \mathbb{E}_{P_{S, X}}[g(S, X)] - \log \mathbb{E}_{P_SP_X}[e^{g(S, X)}]$ consists of linear combinations (expectations) and continuous mappings (logarithm and exponential) of the continuous function $g$, $\Lambda(g)$ is continuous. 
Moreover, $\Lambda(g)$ has a unique optimizer $g^*$.
Therefore, by Lemma~\ref{lem:consistency}, we know that with probability one,
\begin{eqnarray}\label{eq:consistency_2}
|\hat{g}_n(s, x)-\hat{g}_\theta(s, x)| \leq \eta,\; \forall s \in \calS, x \in \calX,
\end{eqnarray}
giving the consistency of the information density estimator.

\subsection{Proof of Theorem~\ref{thm:sample_complexity}}\label{proof_samplecomplexity}
By Hoeffding's inequality \cite{hoeffding1994probability}, for all functions $g$ bounded by $M$, i.e. $|g|\leq M$, we have
\begin{eqnarray}\label{eq:sample_complexity_1}
\begin{aligned}
& \Pr\{ |\mathbb{E}_{P_{S_n, X_n}}[g(S, X)] - \mathbb{E}_{P_{S, X}}[g(S, X)]| > \frac{\eta}{4} \}\leq 2\exp{\left(-\frac{2n^2(\frac{\eta}{2})^2}{(2M)^2n}\right)} = 2\exp{\left(-\frac{n\eta^2}{32M^2}\right)}.
\end{aligned}
\end{eqnarray}
Moreover, since $g_\theta$ is parameterized by $\theta$, we utilize the union bound \cite[Lemma~2.2]{shalev2014understanding} to extend \eqref{eq:sample_complexity_1} for the parameters $\theta$. 
For this purpose, recall that $\Theta \subset \Reals^d$ is compact and bounded by $C$, by the exterior covering number of bounded subspace \cite[pp.~337]{shalev2014understanding}, we know the $r$-covering number $N(r, \Theta)$ of $\Theta$ is upper bounded by 
\begin{eqnarray}\label{eq:covering_number}
N(r, \Theta) \leq \left(\frac{2C\sqrt{d}}{r}\right)^d.
\end{eqnarray}
By \eqref{eq:sample_complexity_1} and \eqref{eq:covering_number}, we have
\begin{eqnarray}
\begin{aligned}
& \Pr\{ \exists \theta_l \in \Theta\; s.t.\; \sup\limits_{g_\theta}|\mathbb{E}_{P_{S_n, X_n}}[g_{\theta_l}(S, X)]- \mathbb{E}_{P_{S, X}}[g_{\theta_l}(S, X)]| > \frac{\eta}{4} \}\leq 2N(r, \Theta)\exp{\left(-\frac{n\eta^2}{32M^2}\right)}.
\end{aligned}
\end{eqnarray}
where $\theta_l$ is in the $r$-cover of $\Theta$. 
Since $\calG(\Theta)$ is compact, we can replace the supremum by maximum.
To make $2N(r, \Theta)\exp{\left(-\frac{n\eta^2}{32M^2}\right)} < \delta$, we have
\begin{eqnarray}
n > \frac{32M^2 (\log N(r, \Theta) + \log\frac{2}{\delta})}{\eta^2}.
\end{eqnarray}
Now, let $r = \frac{\eta}{8L}$, and recall that $g_\theta$ is $L$-Lipschitz continuous with respect to $\theta$, then for any $\theta \in \Theta$, we have with probability one
\begin{eqnarray}
|g_\theta - g_{\theta_l}| \leq L|\theta-\theta_l| \leq L r = L\times\frac{\eta}{8L} = \frac{\eta}{8}.
\end{eqnarray}
By triangular inequality, for any $\theta \in \Theta$, whenever $n > \frac{32M^2 (d\log\frac{16LC\sqrt{d}}{\eta} + \log\frac{2}{\delta})}{\eta^2}$, we have with probability at least $1-\delta$,
\begin{eqnarray}
\begin{aligned}
\max\limits_{g_\theta}|\mathbb{E}_{P_{S_n, X_n}}[g_{\theta}(S, X)] - \mathbb{E}_{P_{S, X}}[g_{\theta}(S, X)]|\leq & \max\limits_{g_\theta}|\mathbb{E}_{P_{S_n, X_n}}[g_{\theta}(S, X)] - \mathbb{E}_{P_{S_n, X_n}}[g_{\theta_l}(S, X)]|\\
& + \max\limits_{g_\theta}|\mathbb{E}_{P_{S_n, X_n}}[g_{\theta_l}(S, X)] - \mathbb{E}_{P_{S, X}}[g_{\theta_l}(S, X)]| \\
& + \max\limits_{g_\theta}|\mathbb{E}_{P_{S, X}}[g_{\theta}(S, X)] - \mathbb{E}_{P_{S, X}}[g_{\theta_l}(S, X)]| \\
\leq & \frac{\eta}{8} + \frac{\eta}{4} + \frac{\eta}{8} = \frac{\eta}{2}
\end{aligned}
\end{eqnarray}
Therefore, we have
\begin{eqnarray}\label{eq:sample_complexity_2}
\begin{aligned}
&\Pr\{ \max\limits_{g_\theta}|\mathbb{E}_{P_{S_n, X_n}}[g_{\theta}(S, X)]- \mathbb{E}_{P_{S, X}}[g_{\theta}(S, X)]| \leq \frac{\eta}{2} \} \geq 1-\delta.
\end{aligned}
\end{eqnarray}
Similarly, starting from 
\begin{eqnarray}
\begin{aligned}
&\Pr\{ \exists \theta_l \in \Theta\; s.t.\; |\log\mathbb{E}_{P_{S_n}P_{X_n}}[e^{g_{\theta_l}(S, X)}]- \log\mathbb{E}_{P_SP_X}[e^{g_{\theta_l}(S, X)}]| \geq \frac{\eta}{4} \}\\
\leq & 2N(r, \Theta)\exp{\left(-\frac{n\eta^2}{32M^2}\right)},
\end{aligned}
\end{eqnarray}
we also conclude that for any $\theta \in \Theta$, whenever $n > \frac{32M^2 (d\log\frac{16LC\sqrt{d}}{\eta} + \log\frac{2}{\delta})}{\eta^2}$, we have with probability at least $1-\delta$,
\begin{eqnarray}\label{eq:sample_complexity_3}
\begin{aligned}
&\Pr\{ \max\limits_{g_\theta}|\log\mathbb{E}_{P_{S_n, X_n}}[E^{g_{\theta}(S, X)}]- \log\mathbb{E}_{P_{S, X}}[e^{g_{\theta}(S, X)}]| \leq \frac{\eta}{2} \} \geq 1-\delta.
\end{aligned}
\end{eqnarray}
Summarizing \eqref{eq:sample_complexity_2} and \eqref{eq:sample_complexity_3}, whenever $n > \frac{32M^2 (d\log\frac{16LC\sqrt{d}}{\eta} + \log\frac{2}{\delta})}{\eta^2}$, for any $\theta \in \Theta$, we have
\begin{eqnarray}
\begin{aligned}
& \Pr\{|\max\Lambda_n(\hat{g}_n(s, x))-\max\Lambda(g(s, x))| \leq \eta\}\\
\geq & \Pr\{\max\limits_{g_\theta}|\mathbb{E}_{P_{S_n, X_n}}[g_{\theta}(S, X)]- \mathbb{E}_{P_{S, X}}[g_{\theta}(S, X)]| +\max\limits_{g_\theta}|\log\mathbb{E}_{P_{S_n, X_n}}[E^{g_{\theta}(S, X)}]\\
& - \log\mathbb{E}_{P_{S, X}}[e^{g_{\theta}(S, X)}]| \leq \eta\}\\
\geq & 1-\delta.
\end{aligned}
\end{eqnarray}
The trimmed information density estimator, in this sense, gives a trimmed (clipped) information density, i.e. $|\hat{g}_n(s, x) -g^*(s, x)| \leq \eta$ if $g^*(s, x) \leq M$ and $|\hat{g}_n(s, x) -g^*(s, x)| \geq \eta$ otherwise.
By the concentration of the information density \cite{polyanskiy2014lecture}, we also know the probability that the information density is clipped is upper bounded, i.e.
\begin{eqnarray}
\Pr\{|g^*(s, x)| \geq M\} \leq e^{-M}.
\end{eqnarray}
Therefore, whenever $n > \frac{32M^2 (d\log\frac{16LC\sqrt{d}}{\eta} + \log\frac{2}{\delta})}{\eta^2}$, for all $s \in \calS$ and $x \in \calX$, we have 
\begin{eqnarray}
\begin{aligned}
&\Pr\{ |\hat{g}_n(s, x) - g^*(s, x)| \leq \eta \}\geq 1-\delta \geq 1- e^{-M},
\end{aligned}
\end{eqnarray}
by choosing $\delta \leq e^{-M}$, and the desire result follows.

\section{Experimental Details}
In this section, we provide detailed experimental setups including architecture of the function $g$ in TIDE, training details for the experiments shown in the main text.

\subsection{GENKI-$4$K Smiling Dataset}
The GENKI-$4$K smiling dataset \cite{genki} contains $2400$ colorful images for training and $600$ for test, where each image, viewed as $X$, is a $64\times 64$ pixels face that is smiling ($S = 1$) or not ($S = 0$).

Since the inputs of the encoder TIDE are images, we use adopt a convolutional neural net with three convolutional layers, two fully-connected layers, and a readout layer.
The convolutional layers have kernels with dimension $(5, 5, 3, 64)$, $(5, 5, 64, 64)$, and $(3, 3, 64, 128)$ respectively.
After flatting the output of the third convolutional layer, we feed the output to two fully-connected layers with $384$ and $192$ neurons respectively.
We train for $100$ epochs using \texttt{AdagradOptimizer} with learning rate $0.0001$ and batch size $256$, and achieve $I(S, X) = 0.594 < H(S) = 1$ bits.

The adversary we used here is also a convolutional neural net with identical structure as the TIDE with the difference that the objective is the cross-entropy loss for classification, and is trained for $150$ epochs using \texttt{AdagradOptimizer} with learning rate $0.005$ and batch size $256$.



\subsection{Celebrity Attributes (CelebA) Dataset}
The CelebA dataset \cite{liu2015faceattributes} contains $202599$ colorful images, where each image is a $218\times 178$ pixels face of a celebrity with $40$ distinct binary labels, including \texttt{smiling}, \texttt{gender}, \texttt{Arched Eyebrows}, etc. 
We select $100000$ face images as $X$ and the private attribute $S$ as smiling or not.

Since the inputs of the encoder TIDE are images, we use adopt a convolutional neural net with five convolutional layers, two fully-connected layers, and a readout layer.
The convolutional layers have kernels with dimension $(5, 5, 3, 64)$, $(5, 5, 64, 64)$, $(3, 3, 128, 128)$, $(3, 3, 128, 128)$, and $(3, 3, 64, 128)$ respectively.
After flatting the output of the third convolutional layer, we feed the output to two fully-connected layers with $384$ and $192$ neurons respectively.
We train for $100$ epochs using \texttt{AdagradOptimizer} with learning rate $0.005$ and batch size $64$, and achieve $I(S, X) = 0.967 \approx H(S) = 1$ bits.

The adversaries we used for emotion and gender detection here are also convolutional neural nets with identical structure as the TIDE with the difference that the objective is the cross-entropy loss for classification.
We train the adversaries for $300$ epochs using \texttt{AdagradOptimizer} with learning rate $0.001$ and batch size $2000$.


\subsection{Politically-Biased Tweets}
We collect $75946$ tweets from more than $20$ online publishers (e.g. CNN, Bloomberg, New York Times) using the Twitter API, and determine its private attribute $S$ as the political bias of being right-wing ($S=0$) and left-wing ($S=1$) according to \cite{tweet_political_bias}.
We clean up the tweets to only keep meaningful terms (i.e. pieces of words), and use bag-of-words representation \cite{manning2010introduction} to tokenize all the pieces of words for each tweet according to term frequency, ending up with $24657$ words ($x_j$).
We order the $x_j$ by the order it appears in the training texts of the Tweets.

The TIDE is a simple feed-forward neural network consists of three hidden layers with \texttt{ReLU} activation with $100$ neurons for each hidden layer, and a readout layer with $32$ neurons.
We train for $50$ epochs using \texttt{AdagradOptimizer} with learning rate $0.005$ and batch size $128$, and achieve $I(S; X) = 0.645$ bits.


\begin{table*}[!tb]
  \footnotesize
  \caption{\footnotesize WMAE of the information density estimation on Gaussian synthetic data ($M = 5$).}
  \label{tab:synthetic}
  \centering
  \begin{tabular}{lclclclclclclclclclclclclcl}
     & \multicolumn{4}{c}{Empirical Plug-In Estimator} & \multicolumn{4}{c}{Kernel Density Estimator} & \multicolumn{4}{c}{TIDE}\\
    \cmidrule(r){2-5}  \cmidrule(r){6-9}  \cmidrule(r){10-13}
    \diagbox{$d$}{$\rho$} & $0.0$ & $0.1$ & $0.2$ & $0.5$ & $0.0$ & $0.1$ & $0.2$ & $0.5$ & $0.0$ & $0.1$ & $0.2$ & $0.5$\\
    \hline
    $1$ & $0.466$ & $0.509$ & $1.092$ & $1.821$ & $0.252$ & $0.434$ & $0.973$ & $1.395$ & \bm{$0.005$} & \bm{$0.007$} & \bm{$0.011$} & \bm{$0.057$} \\
    $10$& $5.305$ & $7.613$ & $9.704$ & $18.245$ & $2.869$ & $4.076$ & $6.698$ & $11.496$ & \bm{$1.010$} & \bm{$1.216$} & \bm{$1.884$} & \bm{$2.503$} \\
  \end{tabular}
\end{table*}

\section{Additional Experiments on Synthetic Data}\label{sec:synthetic}
We apply the TIDE in Section~\ref{sec:estimation} on Gaussian synthetic data to estimate the trimmed information density with limited number of samples and $M = 5$.
We consider two $d$-dimensional multivariate standard Normal random variables $S$ and $X$, with pairwise correlation $\mathrm{corr}(S_i, X_j) = \rho\mathbf{1}_{\{i=j\}}$, $\rho \in (-1, 1)$, $1 \leq i, j, \leq d$. 
Since the KL divergence is invariant to continuous bijective transformations of the considered variables, it is sufficient to consider $S$ and $X$ with standard Normal marginals.
We generate $3000$ samples with $70\%-30\%$ train-test split accordingly.
The TIDE is a simple feed-forward neural network consists of three hidden layers with \texttt{ReLU} activation with $100$, $50$, $50$ neurons for each hidden layer, and a readout layer with $50$ neurons.
We jointly train over the entire training set for $3000$ epochs using \texttt{AdagradOptimizer} with learning rate $0.005$.

We compare the plug-in estimator using empirical distributions (with $30$ bins for quantization), the Gaussian kernel density estimator \cite{bishop2006pattern}, and the TIDE using $3$k samples, and report the Weighted Mean Absolute Error (WMAE) of the information density in Table~\ref{tab:synthetic}, where the weights are chosen as the ground true joint distributions and each number in the table is averaged over $10$ repeated experiments.
Note that since the Normal random variable is continuous, quantized empirical distribution gives loose estimate.
The kernel density estimator performs better than the plug-in estimator but worse than the TIDE due to limited number of samples.

\bibliographystyle{IEEEtran}
\bibliography{aistats2020.bib}

\begin{thebibliography}{10}
\providecommand{\url}[1]{#1}
\csname url@samestyle\endcsname
\providecommand{\newblock}{\relax}
\providecommand{\bibinfo}[2]{#2}
\providecommand{\BIBentrySTDinterwordspacing}{\spaceskip=0pt\relax}
\providecommand{\BIBentryALTinterwordstretchfactor}{4}
\providecommand{\BIBentryALTinterwordspacing}{\spaceskip=\fontdimen2\font plus
\BIBentryALTinterwordstretchfactor\fontdimen3\font minus
  \fontdimen4\font\relax}
\providecommand{\BIBforeignlanguage}[2]{{%
\expandafter\ifx\csname l@#1\endcsname\relax
\typeout{** WARNING: IEEEtran.bst: No hyphenation pattern has been}%
\typeout{** loaded for the language `#1'. Using the pattern for}%
\typeout{** the default language instead.}%
\else
\language=\csname l@#1\endcsname
\fi
#2}}
\providecommand{\BIBdecl}{\relax}
\BIBdecl

\bibitem{bertran2019adversarially}
M.~Bertran, N.~Martinez, A.~Papadaki, Q.~Qiu, M.~Rodrigues, G.~Reeves, and
  G.~Sapiro, ``Adversarially learned representations for information
  obfuscation and inference,'' in \emph{Proc. of International Conference on
  Machine Learning (ICML)}, 2019.

\bibitem{chen2019distributed}
X.~Chen, T.~Navidi, S.~Ermon, and R.~Rajagopal, ``Distributed generation of
  privacy preserving data with user customization,'' \emph{arXiv preprint
  arXiv:1904.09415}, 2019.

\bibitem{huang2018generative}
C.~Huang, P.~Kairouz, X.~Chen, L.~Sankar, and R.~Rajagopal, ``Generative
  adversarial privacy,'' \emph{arXiv preprint arXiv:1807.05306}, 2018.

\bibitem{zemel2013learning}
R.~Zemel, Y.~Wu, K.~Swersky, T.~Pitassi, and C.~Dwork, ``Learning fair
  representations,'' in \emph{International Conference on Machine Learning},
  2013, pp. 325--333.

\bibitem{du2012privacy}
F.~du~Pin~Calmon and N.~Fawaz, ``Privacy against statistical inference,'' in
  \emph{Proc. of IEEE Allerton Conference on Communication, Control, and
  Computing (Allerton)}, 2012, pp. 1401--1408.

\bibitem{Asoodeh_Arimoto}
S.~Asoodeh, M.~Diaz, F.~Alajaji, and T.~Linder, ``Estimation efficiency under
  privacy constraints,'' \emph{IEEE Transactions on Information Theory},
  vol.~65, no.~3, pp. 1512--1534, 2018.

\bibitem{issa2018operational}
I.~Issa, A.~B. Wagner, and S.~Kamath, ``An operational approach to information
  leakage,'' \emph{arXiv preprint arXiv:1807.07878}, 2018.

\bibitem{hsu2018generalizing}
H.~Hsu, S.~Asoodeh, S.~Salamatian, and F.~P. Calmon, ``Generalizing bottleneck
  problems,'' in \emph{Proc. of IEEE International Symposium on Information
  Theory (ISIT)}, 2018.

\bibitem{diaz2018robustness}
M.~Diaz, H.~Wang, F.~P. Calmon, and L.~Sankar, ``On the robustness of
  information-theoretic privacy measures and mechanisms,'' \emph{arXiv preprint
  arXiv:1811.06057}, 2018.

\bibitem{basciftci2016privacy}
Y.~O. Basciftci, Y.~Wang, and P.~Ishwar, ``On privacy-utility tradeoffs for
  constrained data release mechanisms,'' in \emph{2016 Information Theory and
  Applications Workshop (ITA)}.\hskip 1em plus 0.5em minus 0.4em\relax IEEE,
  2016, pp. 1--6.

\bibitem{huang2017context}
C.~Huang, P.~Kairouz, X.~Chen, L.~Sankar, and R.~Rajagopal, ``Context-aware
  generative adversarial privacy,'' \emph{Entropy}, vol.~19, no.~12, p. 656,
  2017.

\bibitem{pinsker}
M.~S. Pinsker, \emph{Information and information stability of random variables
  and processes}.\hskip 1em plus 0.5em minus 0.4em\relax San Francisco:
  Holden-Day, 1964.

\bibitem{han1993approximation}
T.~S. Han and S.~Verd{\'u}, ``Approximation theory of output statistics,''
  \emph{IEEE Transactions on Information Theory}, vol.~39, no.~3, pp. 752--772,
  1993.

\bibitem{evfimievski2003limiting}
A.~Evfimievski, J.~Gehrke, and R.~Srikant, ``Limiting privacy breaches in
  privacy preserving data mining,'' in \emph{Proc. of ACM SIGMOD-SIGACT-SIGART
  symposium on Principles of database systems}, 2003.

\bibitem{ZeroDP}
M.~Bun and T.~Steinke, ``Concentrated differential privacy: Simplifications,
  extensions, and lower bounds,'' in \emph{Theory of Cryptography}.\hskip 1em
  plus 0.5em minus 0.4em\relax Berlin, Heidelberg: Springer Berlin Heidelberg,
  2016, pp. 635--658.

\bibitem{Concentrated_Dwork}
C.~Dwork and G.~N. Rothblum, ``Concentrated differential privacy,'' \emph{arXiv
  preprint arXiv:1603.01887}, 2016.

\bibitem{balle2018improving}
B.~Balle and Y.-X. Wang, ``Improving the {G}aussian mechanism for differential
  privacy: Analytical calibration and optimal denoising,'' in \emph{in Proc. of
  the International Conference on Machine Learning}, 2018.

\bibitem{sarwate2013signal}
A.~D. Sarwate and K.~Chaudhuri, ``Signal processing and machine learning with
  differential privacy: Algorithms and challenges for continuous data,''
  \emph{IEEE signal processing magazine}, vol.~30, no.~5, pp. 86--94, 2013.

\bibitem{chaudhuri2011differentially}
K.~Chaudhuri, C.~Monteleoni, and A.~D. Sarwate, ``Differentially private
  empirical risk minimization,'' \emph{Journal of Machine Learning Research},
  vol.~12, no. Mar, pp. 1069--1109, 2011.

\bibitem{Csiszar67}
I.~Csisz{\'a}r, ``{Information-type measures of difference of probability
  distributions and indirect observations},'' \emph{Studia Sci. Math. Hungar.},
  vol.~2, pp. 299--318, 1967.

\bibitem{valiant2011estimating}
G.~Valiant and P.~Valiant, ``Estimating the unseen: an n/log (n)-sample
  estimator for entropy and support size, shown optimal via new clts,'' in
  \emph{Proc. of ACM symposium on Theory of computing (STOC)}, 2011.

\bibitem{wu2016minimax}
Y.~Wu and P.~Yang, ``Minimax rates of entropy estimation on large alphabets via
  best polynomial approximation,'' \emph{IEEE Transactions on Information
  Theory}, vol.~62, no.~6, pp. 3702--3720, 2016.

\bibitem{gao2017estimating}
W.~Gao, S.~Kannan, S.~Oh, and P.~Viswanath, ``Estimating mutual information for
  discrete-continuous mixtures,'' in \emph{Proc. of Advances in Neural
  Information Processing Systems (NeurIPS)}, 2017, pp. 5986--5997.

\bibitem{vapnik2013nature}
V.~Vapnik, \emph{The nature of statistical learning theory}.\hskip 1em plus
  0.5em minus 0.4em\relax Springer science \& business media, 2013.

\bibitem{belghazi2018mine}
I.~Belghazi, S.~Rajeswar, A.~Baratin, R.~D. Hjelm, and A.~Courville, ``Mine:
  mutual information neural estimation,'' \emph{arXiv preprint
  arXiv:1801.04062}, 2018.

\bibitem{liu2017trimmed}
S.~Liu, A.~Takeda, T.~Suzuki, and K.~Fukumizu, ``Trimmed density ratio
  estimation,'' in \emph{Proc. of Advances in Neural Information Processing
  Systems (NeurIPS)}, 2017.

\bibitem{nguyen2010estimating}
X.~Nguyen, M.~J. Wainwright, and M.~I. Jordan, ``Estimating divergence
  functionals and the likelihood ratio by convex risk minimization,''
  \emph{{IEEE} Transactions on Information Theory}, vol.~56, no.~11, pp.
  5847--5861, 2010.

\bibitem{mcpherson2016defeating}
R.~McPherson, R.~Shokri, and V.~Shmatikov, ``Defeating image obfuscation with
  deep learning,'' \emph{arXiv preprint arXiv:1609.00408}, 2016.

\bibitem{oh2017adversarial}
S.~J. Oh, M.~Fritz, and B.~Schiele, ``Adversarial image perturbation for
  privacy protection a game theory perspective,'' in \emph{2017 IEEE
  International Conference on Computer Vision (ICCV)}.\hskip 1em plus 0.5em
  minus 0.4em\relax IEEE, 2017, pp. 1491--1500.

\bibitem{wu2018towards}
Z.~Wu, Z.~Wang, Z.~Wang, and H.~Jin, ``Towards privacy-preserving visual
  recognition via adversarial training: A pilot study,'' in \emph{Proceedings
  of the European Conference on Computer Vision (ECCV)}, 2018, pp. 606--624.

\bibitem{genki}
T.~{MPL}ab, ``The {MPL}ab {GENKI} {D}atabase, {GENKI-4K} {S}ubset,''
  \url{http://mplab.ucsd.edu}, 2009.

\bibitem{liu2015faceattributes}
Z.~Liu, P.~Luo, X.~Wang, and X.~Tang, ``Deep learning face attributes in the
  wild,'' in \emph{Proc. of International Conference on Computer Vision
  (ICCV)}, 2015.

\bibitem{tweet_political_bias}
A.~Rachez, ``Predicting political bias with python,''
  https://medium.com/linalgo/predict-political-bias-using-python-b8575eedef13,
  2017, accessed: 2019-03-21.

\bibitem{liao2018tunable}
J.~Liao, O.~Kosut, L.~Sankar, and F.~P. Calmon, ``A tunable measure for
  information leakage,'' \emph{arXiv preprint arXiv:1806.03332}, 2018.

\bibitem{sankar2013utility}
L.~Sankar, S.~R. Rajagopalan, and H.~V. Poor, ``Utility-privacy tradeoffs in
  databases: An information-theoretic approach,'' \emph{IEEE Transactions on
  Information Forensics and Security}, vol.~8, no.~6, pp. 838--852, 2013.

\bibitem{makhdoumi2014information}
A.~Makhdoumi, S.~Salamatian, N.~Fawaz, and M.~M{\'e}dard, ``From the
  information bottleneck to the privacy funnel,'' in \emph{Proc. of {IEEE}
  Information Theory Workshop (ITW)}, 2014.

\bibitem{nissim2007smooth}
K.~Nissim, S.~Raskhodnikova, and A.~Smith, ``Smooth sensitivity and sampling in
  private data analysis,'' in \emph{Proceedings of the thirty-ninth annual ACM
  symposium on Theory of computing}.\hskip 1em plus 0.5em minus 0.4em\relax
  ACM, 2007, pp. 75--84.

\bibitem{cuff2016differential}
P.~Cuff and L.~Yu, ``Differential privacy as a mutual information constraint,''
  in \emph{Proc. of the ACM SIGSAC Conference on Computer and Communications
  Security (CCS)}, 2016.

\bibitem{mironov2017renyi}
I.~Mironov, ``R{\'e}nyi differential privacy,'' in \emph{2017 IEEE 30th
  Computer Security Foundations Symposium (CSF)}.\hskip 1em plus 0.5em minus
  0.4em\relax IEEE, 2017, pp. 263--275.

\bibitem{Kifer_Free_Lunch}
\BIBentryALTinterwordspacing
D.~Kifer and A.~Machanavajjhala, ``No free lunch in data privacy,'' in
  \emph{Proceedings of the 2011 ACM SIGMOD International Conference on
  Management of Data}, ser. SIGMOD '11.\hskip 1em plus 0.5em minus 0.4em\relax
  New York, NY, USA: ACM, 2011, pp. 193--204. [Online]. Available:
  \url{http://doi.acm.org/10.1145/1989323.1989345}
\BIBentrySTDinterwordspacing

\bibitem{yamada2011relative}
M.~Yamada, T.~Suzuki, T.~Kanamori, H.~Hachiya, and M.~Sugiyama, ``Relative
  density-ratio estimation for robust distribution comparison,'' in \emph{Proc.
  of Advances in neural information processing systems (NeurIPS)}, 2011.

\bibitem{smola2009relative}
A.~Smola, L.~Song, and C.~H. Teo, ``Relative novelty detection,'' in
  \emph{Proc. of International Conference on Artificial Intelligence and
  Statistics (AISTATS)}, 2009.

\bibitem{sugiyama2007covariate}
M.~Sugiyama, M.~Krauledat, and K.-R. M{\~A}{\v{z}}ller, ``Covariate shift
  adaptation by importance weighted cross validation,'' \emph{Journal of
  Machine Learning Research (JMLR)}, vol.~8, no. May, pp. 985--1005, 2007.

\bibitem{goodfellow2014generative}
I.~Goodfellow, J.~Pouget-Abadie, M.~Mirza, B.~Xu, D.~Warde-Farley, S.~Ozair,
  A.~Courville, and Y.~Bengio, ``Generative adversarial nets,'' in
  \emph{Advances in neural information processing systems}, 2014, pp.
  2672--2680.

\bibitem{sugiyama2012density}
M.~Sugiyama, T.~Suzuki, and T.~Kanamori, \emph{Density ratio estimation in
  machine learning}.\hskip 1em plus 0.5em minus 0.4em\relax Cambridge
  University Press, 2012.

\bibitem{cover2012elements}
T.~M. Cover and J.~A. Thomas, \emph{Elements of information theory}.\hskip 1em
  plus 0.5em minus 0.4em\relax John Wiley \& Sons, 2012.

\bibitem{EGamma}
J.~{Liu}, P.~{Cuff}, and S.~{Verdú}, ``$e_{ {\gamma}}$-resolvability,''
  \emph{IEEE Transactions on Information Theory}, vol.~63, no.~5, pp.
  2629--2658, May 2017.

\bibitem{polyanskiy2010channel}
Y.~Polyanskiy, H.~V. Poor, and S.~Verd{\'u}, ``Channel coding rate in the
  finite blocklength regime,'' \emph{{IEEE} Transactions on Information
  Theory}, vol.~56, no.~5, pp. 2307--2359, 2010.

\bibitem{Barthe}
G.~Barthe and F.~Olmedo, ``Beyond differential privacy: Composition theorems
  and relational logic for f-divergences between probabilistic programs,'' in
  \emph{Automata, Languages, and Programming}.\hskip 1em plus 0.5em minus
  0.4em\relax Berlin, Heidelberg: Springer Berlin Heidelberg, 2013, pp. 49--60.

\bibitem{Balle2019mixing}
B.~Balle, G.~Barthe, M.~Gaboardi, and J.~Geumlek, ``Privacy amplification by
  mixing and diffusion mechanisms,'' \emph{ArXiv}, vol. abs/1905.12264, 2019.

\bibitem{Polyanskiy_Arimoto}
Y.~{Polyanskiy} and S.~{Verd\'u}, ``Arimoto channel coding converse and
  {R\'enyi} divergence,'' in \emph{2010 48th Annual Allerton Conference on
  Communication, Control, and Computing (Allerton)}, Sep. 2010, pp. 1327--1333.

\bibitem{sason2016f}
I.~Sason and S.~Verd\'u, ``$f$-divergence inequalities,'' \emph{IEEE
  Transactions on Information Theory}, vol.~62, no.~11, pp. 5973--6006, 2016.

\bibitem{donsker1983asymptotic}
M.~D. Donsker and S.~S. Varadhan, ``Asymptotic evaluation of certain markov
  process expectations for large time. iv,'' \emph{Communications on Pure and
  Applied Mathematics}, vol.~36, no.~2, pp. 183--212, 1983.

\bibitem{amemiya1985advanced}
T.~Amemiya, ``Asymptotic properties of extremum estimators,'' \emph{Advanced
  econometrics, Harvard university press}, 1985.

\bibitem{newey1994large}
W.~K. Newey and D.~McFadden, ``Large sample estimation and hypothesis
  testing,'' \emph{Handbook of econometrics}, vol.~4, pp. 2111--2245, 1994.

\bibitem{church1990word}
K.~W. Church and P.~Hanks, ``Word association norms, mutual information, and
  lexicography,'' \emph{Computational linguistics}, vol.~16, no.~1, pp. 22--29,
  1990.

\bibitem{alzantot2018generating}
M.~Alzantot, Y.~Sharma, A.~Elgohary, B.-J. Ho, M.~Srivastava, and K.-W. Chang,
  ``Generating natural language adversarial examples,'' \emph{arXiv preprint
  arXiv:1804.07998}, 2018.

\bibitem{manning2010introduction}
C.~Manning, P.~Raghavan, and H.~Sch{\"u}tze, ``Introduction to information
  retrieval,'' \emph{Natural Language Engineering}, vol.~16, no.~1, pp.
  100--103, 2010.

\bibitem{vzliobaite2016using}
I.~{\v{Z}}liobait{\.e} and B.~Custers, ``Using sensitive personal data may be
  necessary for avoiding discrimination in data-driven decision models,''
  \emph{Artificial Intelligence and Law}, vol.~24, no.~2, pp. 183--201, 2016.

\bibitem{van2000empirical}
S.~van~de Geer, \emph{Empirical Processes in M-estimation}.\hskip 1em plus
  0.5em minus 0.4em\relax Cambridge university press, 2000, vol.~6.

\bibitem{hoeffding1994probability}
W.~Hoeffding, ``Probability inequalities for sums of bounded random
  variables,'' in \emph{The Collected Works of Wassily Hoeffding}.\hskip 1em
  plus 0.5em minus 0.4em\relax Springer, 1994, pp. 409--426.

\bibitem{shalev2014understanding}
S.~Shalev-Shwartz and S.~Ben-David, \emph{Understanding machine learning: From
  theory to algorithms}.\hskip 1em plus 0.5em minus 0.4em\relax Cambridge
  university press, 2014.

\bibitem{polyanskiy2014lecture}
Y.~Polyanskiy and Y.~Wu, ``Lecture notes on information theory,'' \emph{Lecture
  Notes for ECE563 (UIUC) and}, vol.~6, pp. 2012--2016, 2014.

\bibitem{bishop2006pattern}
C.~M. Bishop, \emph{Pattern recognition and machine learning}.\hskip 1em plus
  0.5em minus 0.4em\relax springer, 2006.

\end{thebibliography}
\end{document}